%% file: main.tex
\documentclass[nofootinbib
10pt,
nofootinbib,
amsmath,amssymb,
aps,
twocolumn,
pra,
noarxiv,
twopage,
longbibliography
]{revtex4-2}

\usepackage[T1]{fontenc}

\AtBeginDocument{\renewcommand{\selectlanguage}[1]{}}

\usepackage{booktabs,array,makecell,longtable}

\usepackage{graphicx} 
\usepackage[table]{xcolor} 
\usepackage{makecell} 
\usepackage{booktabs} 
\usepackage{makecell} 

\usepackage{amsmath}
\usepackage{mathtools}
\usepackage{amssymb}
\usepackage[mathscr]{euscript} 

\usepackage{amsthm}
\usepackage[colorlinks=true,linkcolor=blue,urlcolor=blue,citecolor=blue,pdfusetitle]{hyperref}

\newcommand\restr[2]{{
  \left.\kern-\nulldelimiterspace 
  #1 
  \vphantom{\big|} 
  \right|_{#2} 
  }}

\newtheoremstyle{break}
  {}
  {}
  {\itshape}
  {}
  {\bfseries}
  {.}
  {\newline}
  {}
\theoremstyle{break}

\theoremstyle{break}
\newtheorem{definition}{Definition}
\newtheorem{proposition}{Proposition}
\usepackage{siunitx}
\usepackage{physics}

\usepackage{tikz}
\usepackage{xcolor} 
\usetikzlibrary{matrix,backgrounds}

\DeclareMathOperator{\diag}{diag}
\DeclareMathOperator{\cnot}{CNOT}

\DeclareMathOperator{\su}{\mathfrak{su}}

\DeclareMathOperator{\Sp}{Sp}

\DeclareMathOperator{\image}{im}

\newcommand{\nodagger}{\vphantom{\dagger}}
\newcommand{\defeq}{\coloneqq}
\newcommand{\eqdef}{\eqqcolon}
\newcommand{\para}{{\mathbin{\|}}}
\newcommand{\eqcomment}[1]{\stackrel{\mathclap{\mbox{$#1$}}}{=}}
\newcommand{\varvec}{\mathbf}
\DeclareMathOperator*{\Span}{span}
\newcommand{\SpanR}[1]{\Span_{#1} \mspace{-1mu} {}_\mathbb{R}}

\newcommand{\galois}{\mathbb{F}_2}

\usepackage{dsfont} 
\usepackage{mathtools}

\begin{document}
	
\title{Entanglement pumping under limited control in solid-state quantum registers}
\author{Daniel L.~Dulog}
\email{daniel.dulog@uni-ulm.de}
\author{Yu Liu}
\email{yu.liu@uni-ulm.de}
\author{Martin B.~Plenio}
\email{martin.plenio@uni-ulm.de}
\affiliation{Institute of Theoretical Physics, Ulm University, Albert-Einstein-Allee 11, 89081 Ulm, Germany}
\affiliation{Center for Integrated Quantum Science and Technology (IQST), 89081 Ulm, Germany}

\date{\today}

\begin{abstract}
    Remote entanglement generation is essential for distributed quantum computing and quantum networks. Entanglement distillation improves fidelity by consuming several low-fidelity pairs to produce fewer high-fidelity ones, but restricted qubit availability and register controllability limit its implementation in solid-state registers. Entanglement pumping reduces this qubit requirement by repeatedly purifying a stored pair with fresh supply pairs. However, its asymptotic fidelity generally saturates below unity and depends on the pumping protocol, while limited control makes only a subset of protocols accessible.
    Here we develop a general framework for analyzing and optimizing entanglement pumping with limited control, using an electron-nuclear register as a representative example. We express the fidelity after each pumping round in terms of the expectation values of the three nontrivial stabilizers of the target Bell state, which we define as stabilizer coordinates. By introducing an additional normalization coordinate, the update of these coordinates from one pumping round to the next can be expressed as a linear transfer map. The eigensystem of the corresponding transfer matrix then determines the asymptotic fidelity and convergence rate.
    We classify pumping protocols according to how the stabilizer coordinates are updated and derive criteria, in terms of the encoding operation and readout string, that determine the corresponding class. Only the biaxial class provides proper pumping for any supply state above the distillation threshold, while other classes can still be useful for particular supply states. For a given supply state, the attainable asymptotic fidelity further depends on how its stabilizer coordinates are assigned to the pumping axes, allowing the available protocols to be compared and optimized systematically. The framework therefore provides a direct way to identify suitable pumping protocols from the operations accessible in a given register.
\end{abstract}

\maketitle
\section{Introduction}
High-fidelity entanglement between spatially separated quantum registers is a central resource for distributed quantum information processing and quantum networks.
In practice, remotely generated entanglement is affected by imperfections during the optical entanglement-generation process and the manipulation and processing on local quantum registers. Entanglement distillation (also called entanglement purification) refers to a probabilistic method that converts several initial noisy entangled pairs into fewer pairs of higher fidelity~\cite{bennett1996, dur2007, horodecki2009}.
The proof-of-principle demonstrations were first realized in linear optics~\cite{kwiat2001, pan2001, yamamoto2003, pan2003}. Later experiments with stationary qubits, including trapped ions~\cite{reichle2006}, color centers~\cite{kalb2017}, and superconducting circuits~\cite{yan2022}, produced distilled pairs that remained available for subsequent operations. Recently, multi-round purification has been demonstrated~\cite{zhou2026}.

A key limitation of entanglement-distillation schemes is their demanding hardware requirements, such as register size and quantum-control efficiency. 
Bennett et al. introduced the BBPSSW protocol in 1996~\cite{bennett1996}, establishing an iterative distillation approach by using multiple noisy pairs. In the same year, Deutsch et al. proposed the extended DEJMPS protocol~\cite{deutsch1996, deutsch1998}. Since then, these schemes have been further extended and generalized in a number of ways~\cite{krastanov2019, rozpedek2018, kalman2025}. In such protocols, the elementary distillation block applies a sequence of local operations, typically Clifford gates, to two noisy pairs, followed by measurements and postselection on one pair. Conditioned on the selected outcomes, the unmeasured pair is probabilistically projected onto a more strongly entangled state, which can then serve as an input pair for the next distillation round. Consequently, reaching an $R$-round distillation requires at least $2^R$ elementary entangled pairs in the ideal binary-tree construction~(see Fig.~\ref{fig:structure} (b)). This exponential pair requirement does not necessitate an equally large register, since the protocol can be implemented in a cascaded manner by storing successful purified pairs in a memory and further distilling them through temporal reordering~\cite{campbell2013}. Nevertheless, the required register size and the number of elementary distillation operations grow linearly and exponentially with the number of distillation rounds, respectively, both remaining challenging for matter-qubit platforms.

Therefore, the entanglement pumping technique illustrated in Fig.~\ref{fig:structure} (a) is preferred, in which each iteration ${\rho_0\otimes\rho_{n-1}\rightarrow\rho_n}$ only requires a fresh supply pair $\rho_0$~\cite{dur1999}. This avoids the storage and swap overhead associated with purified intermediate pairs and keeps the local register size fixed at two qubits. Notably, the supply pair can be replaced by a single carrier qubit passed between the nodes, leaving only one qubit per node~\cite{kim2026}, at the cost of a spin-photon gate at each node. However, for a fixed supply state, the fidelity may saturate at a protocol-dependent value $F_\infty<1$, and no protocol based on local operations and classical communication (LOCC) can be universally beneficial for all entangled inputs~\cite{zang2025}. Thus, the elementary pumping block must be optimized, with the optimal Clifford encoding and measurement depending on the supply state. This issue is particularly relevant for the electron-nuclear spin register, the solid-state system considered, where native weak coupling and the control scheme constrain both the allowed operations and the efficiency of gate implementation. Such registers are found in nitrogen-vacancy~\cite{kalb2017, pompili2021, abobeih2022} and group-IV color centers~\cite{Stas2022, Metsch2019, Beukers2025} in diamond, donors~\cite{Madzik2022, Zhang2026} and T centers~\cite{Bergeron2020, Higginbottom2022} in silicon, rare-earth ions~\cite{Uysal2023}, and carbene molecular spins~\cite{roggors2026}.

In this work, we develop a general framework for entanglement pumping that classifies the pumping dynamics and optimizes control and measurement choices for a given supply state under restricted two-qubit control.
First, as shown in Fig.~\ref{fig:system_pumping}, we consider two registers A and B, each exhibiting a star-shaped configuration in which a central electron spin is hyperfine-coupled to surrounding nuclear spins. One nuclear spin in each register serves as a memory qubit. The electron spins (yellow spins) are driven by modulated microwaves, while the nuclear spins (teal spins) are controlled indirectly through hyperfine coupling. The supply pairs $\rho_0$ are generated between the electron spins via channel $\Gamma$, and the distilled pair is stored in the nuclear spins. Such multi-spin registers are promising building blocks for distributed quantum computing and quantum networks. 
Here, the restrictions come from using zero-mean modulation and retaining only the leading-order terms in the average-Hamiltonian description. Under these assumptions, local nuclear-spin rotations and some two-qubit Clifford gates are excluded. Although such excluded operations can in principle be generated through direct driving or weaker electron-mediated higher-order processes, they generally require longer gate times and additional control sequences~\cite{taminiau2012, casanova2016, casanova2017}. More generally, this trade-off between universal gate control and control efficiency is relevant in realistic systems with finite coherence times.

Second, we introduce a stabilizer-coordinate transfer-matrix description of entanglement pumping. At the $n$th pumping round, we characterize the pumped state using three stabilizer coordinates $(\mu_n,\nu_n,\xi_n)$, corresponding to the expectation values of the three stabilizer operators of the target Bell state. Together with the trace coordinate $t_n$, which tracks the cumulative postselection weight, these form the unnormalized coordinate vector $v_n=(t_n,\mu_n,\nu_n,\xi_n)^T$. These coordinates fully determine the fidelity of the pumped state with respect to the target Bell state. Thus, the coordinates are updated in each pumping step in a linear form $v_{n}=Gv_{n-1}$. The eigenstructure of the transfer matrix $G$ determines the asymptotic pumping fidelity $F_{\infty}$ and convergence rate. 
Based on how the three stabilizer coordinates are updated, we identify two nontrivial pumping types: uniaxial, where one coordinate is updated independently of the other two, and biaxial, where all three are coupled.
We also provide criteria to distinguish the two types.  Unlike uniaxial pumping, biaxial pumping can distill entanglement for any supply state above the distillation threshold. 
Moreover, the transfer matrix $G$ consists of the supply-state coordinates $(\mu_0,\nu_0,\xi_0)$ and fixed numerical entries, whose arrangement is determined by the encoding and measurement. Different arrangements correspond to different assignments of the supply-state coordinates to the signature $(g,b,h)$, leading to different asymptotic fidelities. This allows the optimization to be reduced to comparing the transfer-matrix structures and coordinate assignments of accessible protocols. We numerically find that the optimal assignment satisfies $g\leq b\leq h$. We also consider composite protocols, in which different accessible pumping operations are combined to obtain effective pumping dynamics that are not available from a single protocol. Finally, we apply the framework to practical entanglement-generation schemes, showing how pumping protocols can be selected according to the available register control and the imperfections of the generated states.

Overall, the framework reduces the design of entanglement pumping to identifying the allowed transfer-matrix structure and choosing the appropriate coordinate assignment for a given supply state. It provides a systematic route to efficient distillation under restricted quantum control in the solid-state system. 
This paper is organized as follows. In Sec.~\ref {sec:system_and_control_pump}, we introduce the electron-nuclear register and discuss the controllability under only local control over electron spin. In Sec.~\ref{sec:entanglement_distillation}, we present the framework of entanglement pumping, identifying the pumping type and selecting the optimal protocol. Sec.~\ref{sec:entanglement_generation} discusses the implementation within entanglement generation, and Sec.~\ref{sec:conclusion} concludes with an outlook.

\section{System \& Controllability}
\label{sec:system_and_control_pump}
\begin{figure}
    \centering
    \includegraphics[width=0.95\linewidth]{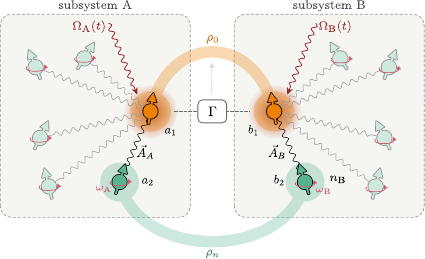}
    \caption{The bipartite system for entanglement distillation. The electron spins $a_1, b_1$ (yellow) are controllable via control fields $\Omega_A, \Omega_B$, and a process $\Gamma$ generates an entangled state $\rho_0$ between the two electron spins. For each subsystem, a nuclear spin $a_2, b_2$ is selected to host the target entangled pair of the distillation protocol.}
    \label{fig:system_pumping}
\end{figure}

In this section, we determine which local operations are available within a single register, since they fix the encoding gates $U = U_A U_B$ from which pumping protocols can be constructed. We consider two spatially separated registers $A$ and $B$, each consisting of a central electron spin that is hyperfine-coupled to surrounding nuclear spins (Fig.~\ref{fig:system_pumping}). For now, we focus on the control properties of the registers alone and disregard the mechanism of entanglement generation, which will be further discussed in Sec.~\ref{sec:entanglement_generation}. 

As an exemplary system with limited control properties, we consider registers constituted by the electron spin of a nitrogen-vacancy (NV) center coupled to multiple nuclear spins via dipolar coupling. Interactions between nuclear spins are neglected, assuming them to be either negligible on the experimental timescale or actively suppressed. Thus, such a register admits a star-connected geometry, giving rise to an interesting control problem (see Fig.~\ref{fig:system_pumping}). Given the setup of this system, we have two classes of qubits within this register: the electron spins are controllable and provide the option to measure and initialize their state, while the nuclear spins have none of these capabilities and their evolution can only be indirectly mediated by the electron spins. Together with the star-shaped connectivity, we uncover the implication on the controllability of the entire register.
For a single partition of the system ($A$ or $B$), we describe the system by the Hamiltonian $(\hbar = 1)$
\begin{align}
    H(t) &= D S_z^2 - \gamma_e \vec{B} \cdot \vec{S} \nonumber \\
    &\phantom{=}- \gamma_n \sum_j \vec{B} \cdot \vec{I}_j +  \vec{S} \cdot \sum_j \varvec{A}_j \cdot \vec{I}_j + \vec{\Omega}(t) \cdot \vec{S},
\end{align}
where $D$ is the zero-field splitting of the nitrogen-vacancy center, $\vec{B}$ is the external static magnetic field, $\varvec{A}_j$ is the hyperfine tensor of the $j$-th nuclear spin, $\gamma_e, \gamma_n$ are the gyromagnetic ratios of the electron spin and the nuclear spins, and $\vec{\Omega}(t)$ is the magnetic control field resonant with the electron spin. We furthermore assume perfect alignment of the magnetic field $\vec{B} = B \hat{e}_z$. Selecting system parameters so that the electronic transitions $D \pm \gamma_e B$ dominate the hyperfine couplings allows a simplification of the Hamiltonian, because all elements $(\varvec{A}_j)_{ab}$ with $a \neq z$ can be neglected because the interaction is unable to induce electronic transitions (secular approximation). Moreover, let the control term $\vec{\Omega}(t) \cdot \vec{S}$ be resonant with the transition between one of the excited states $m_s = \pm 1$ and $m_s = 0$. If the control field $\vec\Omega(t)$ is resonant with the transition $|0\rangle\leftrightarrow|m_s\rangle$ for a fixed $m_s\in\{+1,-1\}$, and its Rabi frequency is small compared to the detuning $2\abs{\gamma_e B}$ of the other transition $|0\rangle\leftrightarrow|{-m_s}\rangle$, then an electron spin initialized in $|0\rangle$ never populates $|{-m_s}\rangle$, and we can restrict the spin-1 system to the effective two-level subspace $\mathrm{span}\{|0\rangle,|m_s\rangle\}$,
\begin{align}
\label{eq:hamiltonian_secular_and_qubit_approx}
H'(t) = - \sum_j \vec{\omega}_j \cdot \vec{I}_j + \frac{m_s}{2}  \sigma_z\sum_j \vec{A}_j \cdot \vec{I}_j + \vec{\Omega}(t) \cdot \vec{\sigma},
\end{align}
with $\sigma_x, \sigma_y, \sigma_z$ the spin operators of the effective spin-$\tfrac{1}{2}$ approximation of the electron spin and $\vec{A}_j$ the hyperfine vector emerging from $\varvec{A}_j$ after the secular approximation. In the interaction picture $U_I(t)$ with respect to the effective Larmor precession $\vec{\omega}_j$ of the nuclear spins and the system control $\vec{\Omega}(t)$ defined by
\begin{align}
    \pdv{t} U_I(t) = i U_I(t) \left[ \vec{\Omega}(t) \cdot \vec{\sigma} - \sum_j \vec{\omega}_j \cdot \vec{I}_j \right],
\end{align}
we find the effective Hamiltonian
\begin{align}
\label{eq:control_hamiltonian_pump}
    H_I(t) = (\vec{f}(t) \cdot &\vec{\sigma}) \sum_j \left[ g_j\cos(\omega_j t) \hat{x}_j \right. \nonumber\\ 
    & \left. + g_j\sin(\omega_j t) \hat{y}_j + h_j \hat{z}_j \right] \cdot \vec{I}_j.
\end{align}
Here $\hat{x}_j, \hat{y}_j, \hat{z}_j$ are the normalized vectors of the coordinate systems $\vec{x}_j \propto \vec{A}_j - (\vec{A}_j \cdot \hat{\omega}_j) \hat{\omega}_j$, $\vec{y}_j = \hat{\omega}_j \times \vec{A}_j$ and $\hat{z}_j = \hat{x}_j \times \hat{y}_j$ with $\hat{\omega}_j = \vec{\omega}_j / \abs{\vec{\omega}_j}$. The prefactors $g_j, h_j$ arise from the transformation to the interaction picture. We call the function $\vec{f}(t)$ defined by $(\vec{f}(t) \cdot \vec{\sigma}) \defeq U^{\phantom{\dagger}}_I \sigma_z U_I^\dagger$ the generalized modulation function.

To analyze the set of gates that can be constructed from such a control problem, we use the concept of the Dynamical Lie Algebra (DLA) that forms a cornerstone of quantum control theory \cite{dalessandro2022}. We continue by introducing the definitions.
\begin{definition}[Lie closure]
    Let  $X$ be a set of linearly independent vectors and $[\, \cdot\,, \,\cdot \,]$ a Lie bracket (for matrix Lie algebras, a commutator).
    The \textbf{Lie closure} of a set X is denoted by the bracket $\langle X \rangle_\text{Lie}$ and forms the vector space spanned by the maximal linearly independent set of repeated commutators $[X_n, \ldots, [X_2, X_1]], \; X_1, \ldots, X_n \in X$.
\end{definition}
\begin{definition}[Dynamical Lie Algebra]
Let $X(t)$ be the solution of a control problem
\begin{align}
    \dot{X}(t) = -i H(f(t)) X, \quad X(0) = X_0,
\end{align}
where $f \in \mathcal{C}$ is a function that models the control of the system that is restricted to the function space $\mathcal{C}$. The \textbf{dynamical Lie algebra} $\mathfrak{g}$ associated with this system is defined by
\begin{align}
    \mathfrak{g} \defeq \langle -i \, H(f) \mid f \in \mathcal{C} \rangle_\text{Lie}.
\end{align}
\end{definition}

Once we have determined the set of admissible controls in our system, we can directly determine the gate set, because the group of unitary operations that are possible with the control $\mathcal{C}$ can be expressed as
\begin{align}
    e^\mathfrak{g} \defeq \{e^{-iH} \mid -i H \in \mathfrak{g}\}.
\end{align}
In our case, the set of admissible control functions $\mathcal{C}$ are maps to the set of unit vectors $\vec{f}(t): \mathbb{R} \rightarrow S^2 \subset \mathbb{R}^3$. Then, the requirements on the Rabi frequency $\vec \Omega (t)$ translate to further restrictions on the control properties of the system. For example, a bound on the Rabi frequency $\abs*{\vec{\Omega}(t)} \leq \Omega_\text{max}$ would then limit the set of control functions to functions $\vec{f}$ where the derivative is bounded by a constant $\abs*{\dv*{\smash{\vec{f}}}{t}} \leq C \equiv \text{const} \,\forall t$. Another common constraint imposed on dynamical decoupling sequences is that the modulation function must have a zero mean, $\langle \vec{f}(t) \rangle_T = 0$, over a single period $T$. This restriction suppresses control-field detuning terms $\Delta S_z$, rendering the control sequence robust against such errors. The trade-off for this robustness is the disappearance of the constant terms $h_j \hat{z}_j$ in Eq.~\eqref{eq:control_hamiltonian_pump}. We denote this set of control functions by $\mathcal{C}_0$ and find the associated DLA $\mathfrak{g}_0$ for a two spin system:
\begin{align}
    \mathfrak{g}_0 = -i \SpanR{} \{& XI, YI, ZI, IZ, \nonumber \\
    &XX, XY, YX, YY, ZX, ZY\}.
\end{align}
We note that this DLA is valid only to the first order in the Magnus expansion (i.e.~in the validity regime of average Hamiltonian theory). Although the vanishing time-averaged term $\langle A_\parallel ZZ \rangle_T = 0$ can restore full controllability at higher orders \cite{taminiau2014a}, these higher-order contributions typically exhibit smaller operator norms, resulting in slower gate operations suppressed by a factor of $\sim A_\parallel / \omega_j$. Consequently, our approximation to first order can be seen as an effective restriction to the fastest gates available in this system. The resulting dynamical Lie algebra is generated by $\mathfrak{g}_0 = \langle XI, YI, ZX, ZY \rangle_\text{Lie}$, ensuring that arbitrary gates in $e^{\mathfrak{g}_0}$ can be efficiently compiled from high-fidelity operations performed through adaptive XY (AXY) pulse sequences \cite{dulog2026, casanova2015a}.

Evidently, with this approximation the DLA $\mathfrak{g}_0$ does not provide universal control $\mathfrak{g}_0 \subsetneq \su(4)$, since all terms of the form $AZ$ with $A \in \{X,Y,Z\}$ are excluded from $\mathfrak{g}_0$. We will use the gate set $e^{\mathfrak{g}_0}$ as an example for a restricted control system and demonstrate that entanglement distillation is possible even with control restrictions. We also note that it is possible to establish full controllability on a centrally controlled system by composition of first-order gates, which we will discuss in Appendix~\ref{sec:full_controllability_first_order}. 

\section{Description of entanglement distillation}
\label{sec:entanglement_distillation}
\begin{figure*}
    \centering
    \includegraphics[width=0.95\linewidth]{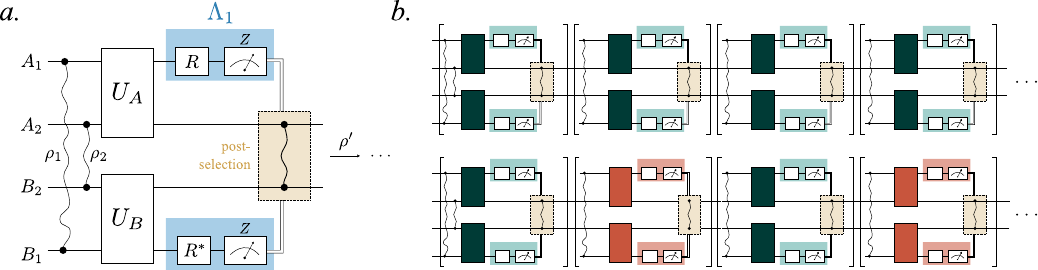}
    \caption{Circuit representations of the distillation protocol. (a) A single distillation step with protocol configuration $\mathscr{P}(U, \Lambda_1)$ with $U = U_A U_B = U_A \tau(U_A)^*$ and $R$ a local rotation that fixes $\Lambda_1$ into one of the three matched strings. (b) Sequence of distillation steps in a homogeneous (top) and composite (bottom) configuration. While the homogeneous protocol has identical iterations, composite protocols alternate steps between two or more different configurations $\mathscr{P}(U_i, \Lambda_1^{(i)})$ indicated by green and red components.} 
    \label{fig:entanglement_protocols}
\end{figure*}

Entanglement distillation utilizes local quantum operations and classical communication (LOCC) to probabilistically convert a larger ensemble of noisy, mixed entangled pairs into a smaller set of pairs with higher fidelity. The primary figure of merit is the entanglement fidelity $F(\rho)$, defined as the overlap of the state $\rho$ with a specific target Bell state $|\chi\rangle$:
\begin{align}
    F(\rho) \defeq \mel{\chi}{\rho}{\chi}.
\end{align}
Although this figure of merit is not an entanglement measure in the strict sense \cite{plenio2007}, it is a practical, directly accessible metric for how closely a distilled state approaches the target Bell state. In a typical distillation step, the entangled pairs are distributed between two spatially separated parties, Alice ($A$) and Bob ($B$). Every entanglement distillation protocol proceeds in three steps (see also Fig.~\ref{fig:entanglement_protocols}(a) for visualization): 
\begin{enumerate}
    \item \textbf{Local encoding.} Alice and Bob apply a local joint unitary gate $U = U_A U_B$ across their respective qubits. To preserve the symmetry of the shared state, we focus on symmetric gates of the form $$U = U_A U_B \quad \text{with} \quad U_A = \tau(U_B)^*,$$ where $\tau$ is the mirror operator that exchanges Alice's and Bob's respective subsystems (see Appendix~\ref{sec:stabilizer_coordinates} for details).
    \item \textbf{Measurement.} Measurements are performed on a subset of the pairs (the auxiliary pairs).
    \item \textbf{Post-selection.} Based on the classical measurement outcomes, the remaining pairs are either retained or discarded according to a predefined post-selection rule.
\end{enumerate}
The simplest setup for entanglement distillation protocols is the conversion of two mixed entangled two-qubit pairs $\rho_1, \rho_2$ into a single pair $\rho_1 \otimes \rho_2 \rightarrow \rho'$ with ideally $F(\rho') > F(\rho_1), F(\rho_2)$. We can distinguish two types of distillation schemes by the input state $\rho_1 \otimes \rho_2$. The first type uses identical input states $\rho \defeq \rho_1 = \rho_2$ which offers advantageous convergence properties to the desired target state. The two textbook representatives of this type are the previously mentioned BBPSSW and DEJMPS protocols. For the latter, one can prove that after $n$ iterations, the output state converges to a perfect Bell state $F(\rho_n) \stackrel{n \rightarrow \infty}{\longrightarrow} 1$ for all initial states satisfying $F(\rho_0) > \frac{1}{2}$ \cite{macchiavello1998}. We will call this setup a \textit{recursive protocol}, since every iteration of the distillation protocol depends only on the output state of the previous iteration $\rho_n \otimes \rho_n \rightarrow \rho_{n+1}$.

\begin{figure}
    \centering
    \includegraphics[width=0.95\linewidth]{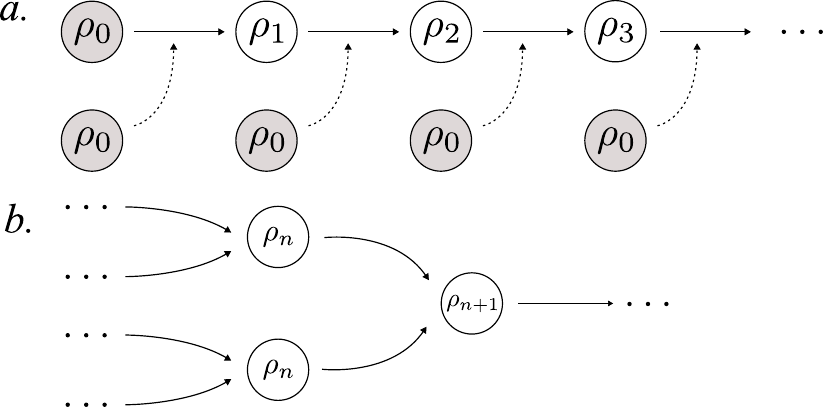}
    \caption{Entanglement distillation protocol structures for a two-pair input. (a) The entanglement pumping scheme, where the initially available entangled pair $\rho_0$ is supplied at each step. (b) The distillation protocol with recursive structure, where the input pairs result from the previous iteration.}
    \label{fig:structure}
\end{figure}

When there is only one qubit in the system that has readout capability, as is the case in our solid-state quantum register, the recursive method amounts to many swap operations within the registers which generate an enormous experimental overhead. Furthermore, in order to create an output state $\rho_n$, one needs two $(=2^1)$ instances of $\rho_{n-1}$, four $(=2^2)$ instances of $\rho_{n-2}$, and hence $2^n$ instances of the initial state $\rho_0$. This imposes a massive constraint on the size of the quantum register, or conversely, the number of possible protocol repetitions. Since solid-state registers are typically constrained in both size and connectivity, we focus on the entanglement pumping procedures in this work. Suppose that an initial, imperfect entangled pair $\rho_0$ with fidelity $\tfrac{1}{2} \leq F_0 \defeq F(\rho_0) < 1$ can be generated between two parties $A$ and $B$. In \textit{entanglement pumping} protocols, one supplies the initial pair $\rho_0$ at every iteration, so that the distillation step is $\rho_0 \otimes \rho_n \rightarrow \rho_{n+1}$. In Fig.~\ref{fig:structure}, the two different approaches to entanglement distillation are illustrated.

Entanglement pumping is a weaker form of distillation, as the fidelity $F_n$ after step $n$ does not generally converge to a perfect Bell pair and saturates at a limiting value $F_\infty <1$ rather than approaching unity. Despite this deficiency, pumping algorithms require only a fixed number of qubits (in our case only two qubits per node) and therefore are easily executed on connectivity- or controllability-limited systems. We analyze entanglement pumping by introducing a stabilizer coordinate formalism, which reduces the distillation dynamics to an eigenvalue problem of a linear map. Afterward, we classify the permissible protocols by their distillation properties and identify optimal strategies tailored to a given supply state $\rho_0$.

\subsection{Stabilizer coordinates}
\label{sec:definition_stabilizer_coordinates}
Entanglement distillation and quantum error correction are conceptually very similar, because in both cases one tries to improve a state by expanding the state space, connects the expanded spaces by an encoding operation $U$, performs measurements on the auxiliary space and manipulates the state according to the measurement outcome. In fact, as soon as the idea of entanglement pumping protocols was born, connections to quantum error correction were established \cite{bennett1996b, krastanov2019, dur2007}. Thus, it appears reasonable to draw on the toolkit of quantum error correction when describing distillation protocols. A useful concept in quantum error correction is the so-called stabilizer formalism \cite{nielsen2010, gaitan2008, calderbank1998, gottesman1996, gottesman1997}. We will proceed by briefly introducing the stabilizer formalism and then derive our new framework, the stabilizer coordinates, from them. The stabilizer coordinates constitute an efficient parametrization of the problem and discard all irrelevant information from the initial state $\rho_0$. Moreover, we can reduce a distillation step to a linear equation, so that studying the dynamics of the pumping protocol can be reduced to an eigenvalue problem. The main performance metrics of a protocol, namely the limit fidelity $F_\infty$ and the convergence speed $\kappa^{-1}$ towards $F_\infty$ can then be easily determined by calculating the eigensystem of the linear map.

Quantum states and subspaces of a Hilbert space are elegantly described by the so-called stabilizer formalism. For a finite dimensional system, instead of specifying a state through a vector $\ket{\psi} \in \mathbb{C}^n$ and evolving it via $U(t) \ket{\psi}$, we represent a state as the common $+1$-eigenspace of a set of mutually commuting operators from the Pauli group $S =\langle A_1, \ldots, A_n \rangle = \{  \prod_{j=1}^nA_j^{k_j} \mid k_1, \ldots, k_n \in \{{0,1 }\}\}$, i.e.
\begin{align}
    A \ket{\psi} = \ket{\psi} \quad \forall A \in S.
\end{align}
Then the evolution of the state can be tracked by evolving the generators $U A_j U^\dagger$. As an example, take the state $\tfrac{1}{\sqrt{2}}(\ket{000} + \ket{111})$, which is stabilized by the operators $Z_1 Z_2$, $Z_2 Z_3$ and $X_1 X_2 X_3$. Since $$Z_m Z_n \ket{k_1 k_2 k_3 } = (-1)^{k_m + k_n} \ket{k_1 k_2 k_3},$$ the space stabilized by stabilizers $Z_1Z_2$ and $Z_2Z_3$ is spanned by $\ket{000}$ and $\ket{111}$. The stabilizer $X_1 X_2 X_3$ then fixes a specific state from this subspace, which is its $+1$-eigenvector $\tfrac{1}{\sqrt{2}}(\ket{000} + \ket{111})$.

Let us now introduce a few notational conventions. The qubits of the $j$-th pair owned by Alice and Bob are denoted by $a_j$ and $b_j$, respectively. The state $\rho_0$ describes the entangled pair supplied at every iteration of the protocol, and $\rho_n$ is the state obtained after the $n$-th iteration with $\rho_0 \otimes \rho_{n-1} \rightarrow \rho_n$. Moreover, we will introduce this formalism for the simplest case of two entangled pairs and use the convention that the first pair is measured and the second pair is purified.

\begin{figure}
    \centering
    \includegraphics[width=0.8\linewidth]{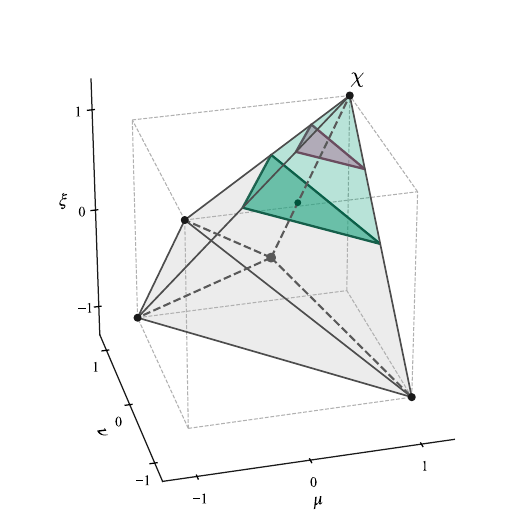}
    \caption{The state space in the stabilizer coordinates is a tetrahedron. The point $(1,1,1)^T$ represents the optimization target $\dyad{\chi}{\chi}$ and $(0,0,0)^T$ the maximally mixed state. The green 2-simplex (triangle) defines the boundary of the subspace of states satisfying $F \geq \tfrac{1}{2}$ (light green). A simplex parallel to it (purple) defines the subset of iso-fidelity states.}
    \label{fig:tetrahedron}
\end{figure}

A Bell state $\ket{\chi}$ has two stabilizers $M$ and $N$, given by
\begin{align}
    M = s_M^\chi X_1 X_2, \quad N = s_N^\chi Z_1 Z_2,
\end{align}
for signs $(s_M^\chi, s_N^\chi) \in \{\pm1\}^{\times 2}$ depending on the specific Bell state $\ket{\chi}$. The stabilizer group is then $S = \langle I, M, N \rangle = \{M, N, MN\}$ and the projector onto the stabilized state is
\begin{align}
    \dyad{\chi}{\chi} &= \frac{1}{2}(I + M) \; \frac{1}{2} (I + N) \nonumber \\ &= \frac{1}{4} (I + M + N + MN).
\end{align}
With this, we immediately see that the fidelity $F(\rho)$ only depends on the expectation values $\mu \defeq \expval{M}$, $\nu \defeq \expval{N}$, $\xi = \expval{MN}$ because
\begin{align}
    F(\rho) &= \mel{\chi}{\rho}{\chi} = \Tr(\dyad{\chi}{\chi} \rho) \nonumber \\
    &= \frac{1}{4} (1 + \Tr(M \rho) + \Tr(N \rho) + \Tr(MN \rho)) \nonumber \\
    &= \frac{1}{4} (1 + \expval{M} + \expval{N} + \expval{MN})  \nonumber \\
    &= \frac{1}{4} (1 + \mu + \nu + \xi).
\end{align}
Thus, we associate a given entangled pair $\rho$ with a point in the coordinate system $(\mu, \nu, \xi)$ and call them the \textit{stabilizer coordinates}. All coordinates take values from $-1$ to $+1$ and the space of physical states is defined by a tetrahedron $\mathscr{T}$ (see Fig.~\ref{fig:tetrahedron}). Each face of the tetrahedron is defined by the positivity of the populations $p_i \geq 0$ and the vertices represent the four Bell states, while $(1,1,1)^T$ embodies the target Bell state $\dyad{\chi}{\chi}$.

As we will see, the evaluation of the distillation properties of a given protocol becomes easily accessible in this coordinate system. Let us denote the local entangling gate of the distillation protocol by $U = U_A U_B$. Moreover, the observable $\widetilde{O} = U^\dagger O U$ denotes the conjugate of the observable $O$. A post-selection of the measurement result of pair $1$ can be described by a projector $\Pi_1$. As an example, the typical coincidental outcome rule is described by the projector
\begin{align}
    \Pi_1 &= \dyad{0_{a_1}0_{b_1}}{0_{a_1} 0_{b_1}} + \dyad{1_{a_1}1_{b_1}}{1_{a_1} 1_{b_1}} \nonumber \\
    &= \frac{1}{2} ( I + Z_{a_1} Z_{b_1}) \eqdef \frac{1}{2} (I + \Lambda_1).
\end{align}
Here, $\Lambda_1$ can be seen as a parity check operator on the measured pair $1$. In our case, we won't fix the post-selection to a specific strategy and treat the projection $\Pi_1$ and the associated Pauli string $\Lambda_1 \in \{XX, -YY, ZZ\}$ as part of the protocol specification $\mathscr{P}(U, \Lambda_1)$. With these definitions, the success probability is given by $p = \Tr(\Pi_1 U\rho U^\dagger) = \tfrac{1}{2}(1 + \expval*{\tilde{\Lambda}_1})$ and the expectation value of an operator $A_2$ acting on pair $2$ (so $\Lambda_1$ and $A_2$ have disjoint support and thus $[A_2, \Pi_1]=0$) is
\begin{align}
    \expval{A_2}_\Pi = \frac{\Tr{A_2 \Pi_1 U \rho U^\dagger \Pi_1}}{\Tr{\Pi_1 U \rho U^\dagger}}.
\end{align}
For a stabilizer on the second pair $A_2 \in \{M, N, MN\}$, $\langle A_2 \rangle_\Pi$ is the value of the coordinates of the retained pair after an iteration of the protocol. Due to the commutation of the observable and the projector $[A_2, \Pi_1] = 0$ and the cyclic property of the trace, we can rewrite the numerator to $\Tr(A_2 \Pi_1 U \rho U^\dagger \Pi_1) = \Tr(U^\dagger A_2 \Pi_1 U \rho) = \Tr(U^\dagger A_2 \Pi_1 U \rho) = \expval*{\widetilde{A}_2 \widetilde{\Pi}_1} = \tfrac{1}{2} ( \expval*{\widetilde{A}_2} + \expval*{\widetilde{A}_2 \widetilde{\Lambda}_1})$. A similar calculation for the denominator leads to the formula
\begin{align}
    \label{eq:stabilizer_update_rule}
    \expval{A_2}_\Pi = \frac{\expval*{\widetilde{A}_2} + \expval*{\widetilde{A}_2 \widetilde{\Lambda}_1}}{1 + \expval*{ \widetilde{\Lambda}_1}}.
\end{align}
This expression constitutes the update formula of a distillation protocol $\mathscr{P}(U, \Lambda_1)$ for the stabilizer coordinates $(\mu, \nu, \xi)$. We regard $\Lambda_1$ as part of the protocol description, because in the following, we will allow a coincident change of the measurement basis, i.e.~$\Lambda_1 \in \{XX, -YY, ZZ\}$.

The coordinate update rule according to Eq.~\eqref{eq:stabilizer_update_rule} is easy to evaluate because one only needs to track the evolution of the matched Pauli strings under the gate $U$. Furthermore, for mirrored Clifford gates $U = U_A U_B$, the update formula is closed in terms of the coordinates, i.e., $(\mu_{n+1}, \nu_{n+1}, \xi_{n+1}) = f(\mu_n, \nu_n, \xi_n)$, so $(\mu, \nu, \xi)$ can be rightfully called a coordinate system (see Appendix~\ref{sec:stabilizer_coordinates}). Obviously, the relation between a set of stabilizer coordinates and a pair $\rho$ is not one-to-one. Since coordinates and populations are related by linear equations (see Eq.~\eqref{eq:populations_expressed_in_coords}), we can define the equivalence relation
\begin{align}
    \rho \sim \rho' \quad :\Longleftrightarrow \quad \diag(\rho)_\text{Bell} = \diag(\rho')_\text{Bell},
\end{align}
so two entangled pairs are equivalent in terms of mirrored distillation protocols if the Bell populations of the two states agree.

For analyzing convergence properties of a distillation protocol, the non-linearity of the update rule in Eq.~\eqref{eq:stabilizer_update_rule} complicates the matter. This issue can be circumvented by introducing another coordinate
\begin{align}
    t \defeq \langle I \rangle = \Tr(\rho), \quad \mu \defeq \langle M \rangle, \quad \nu \defeq \langle N \rangle, \quad \xi \defeq \langle MN \rangle,
\end{align}
so that an entangled pair is described by a four-vector $v = (t, \mu, \nu, \xi)$, and simultaneously removing the normalization from the update rule Eq.~\eqref{eq:stabilizer_update_rule}
\begin{align}
\label{eq:update_rule_linear}
    \expval{A_2}_\Pi = \frac{1}{2}\left( \expval*{\widetilde{A}_2} + \expval*{\widetilde{A}_2 \widetilde{\Lambda}_1} \right).
\end{align}
Since this equation is linear in the stabilizer coordinates, we can rewrite it in terms of a matrix $G$, which we will call the \textit{transfer matrix} inducing the \textit{transfer map}
\begin{align}
    v_{n+1} = G v_n,
\end{align}
where $v_n$ is the coordinate vector after protocol iteration $n$. Kim \textit{et al.}~\cite{kim2026} recently formulated an equivalent linearization for Bell populations in carrier-assisted entanglement pumping. Note that the elements of $G = G(v_0)$ are the stabilizer coordinates $v_0$ of the supply state $\rho_0$, and their position within the matrix is governed by the encoding $U$. Without normalization, the coordinate vector $( \, \cdot \,, \mu, \nu, \xi)$ corresponds to an unphysical, non-normalized state, but the additional coordinate $t = \langle I \rangle = \Tr(\rho) \geq 0$  tracks the trace so that the physical coordinates can be recovered immediately by dividing the vector by $t$. We now demonstrate that the asymptotic dynamics of the distillation process converges to a fixed point $F_\infty$ set by the leading eigenvector $v_\star = (t_\star, \mu_\star, \nu_\star, \xi_\star)$ of $G$. The asymptotic fidelity of the entanglement distillation protocol is thus
\begin{align}
    F_\infty = \frac{1}{4} \left(1 + \frac{\mu_\star + \nu_\star + \xi_\star}{t_\star} \right).
\end{align}
Assume that the supply state $v_0$ satisfies the distillation condition $F_0 > \tfrac{1}{2} \Longrightarrow \mu_0, \nu_0, \xi_0 > 0$, which allows us to infer that all elements of $G$ are non-negative, because each element of $G$ is either an entry of $v_0$ or zero. Furthermore, we assume that $G$ is a primitive matrix, i.e.~$\exists k \in \mathbb{N}_+ : \; G^k >0$. Let $G v_i = \lambda_i v_i$ be the eigenvalue equation of $G$ with ordering $\abs{\lambda_1} \geq \abs{\lambda_2} \geq \ldots \geq \abs{\lambda_m}$. By the Perron-Frobenius theorem~\cite{horn2012}, a primitive matrix is guaranteed to have a dominant eigenvalue that is strictly greater than any other eigenvalue $\abs{\lambda_\star} \defeq \abs{\lambda_1} > \abs{\lambda_2} \geq \ldots \geq \abs{\lambda_m}$. $\lambda_\star$ is algebraically simple and the associated leading eigenvector $v_\star \defeq v_1$ (Perron-Frobenius eigenvector) is the fixed point to which the long-term dynamics converges. To show this, expand a generic vector in the eigenbasis of $G$, $w = \sum_{l=1}^m c_l v_l$ and calculate the action of $G^k$ on $w$ with the property that $\lambda_l / \lambda_\star <1 \; \forall l \neq 1$ (von-Mises iteration \cite{mises1929})
\begin{align}
\frac{1}{\lambda_\star^k} G^k w = \left[ c_1 v_\star + \sum_{l=2}^m c_l \left( \frac{\lambda_l}{\lambda_\star} \right)^k v_l \right] \xrightarrow{k \rightarrow \infty} c_1 v_\star.
\end{align}
Moreover, the ratio $\kappa \defeq \lambda_2 / \lambda_\star$ governs the convergence speed towards the fixed point. The prefactor $c_1$ is irrelevant, as the fidelity of a four-vector is invariant under scalar multiplication $F(\alpha v) = F(v)$. Recasting the asymptotic dynamics of a distillation protocol as a transfer-matrix eigenvalue problem substantially streamlines its analysis: it enables systematic comparison across protocols without resorting to computationally demanding direct simulation and exposes dynamical equivalences between different input states $\rho_1 \otimes \rho_2$ and gates $U$ that eliminate irrelevant degrees of freedom from the outset.

Anticipating the classification of entanglement pumping protocols presented later (Sec.~\ref{sec:classification_of_pumping_protocols}), we observe that the relevant transfer matrices take the form (with $\alpha, \beta, \gamma \in \{ \mu, \nu, \xi\}$)
\begin{align}
    G &\equiv G_{\delta \leftarrow g \alpha}(g, b, h) \nonumber \\
      &\defeq \frac{1}{2} P_{\alpha \delta} \left[ M(1, g)_{(t, \alpha)} \oplus M(b,h)_{(\beta, \gamma)} \right],
      \label{eq:transfer_matrix_structure} \\
      &\phantom{{}\defeq{}} \text{with} \quad
        M(x,y) = \begin{pmatrix} x & y \\ y & x \end{pmatrix}, \nonumber
\end{align}
where the indices $(a,b)$ denote the subspace $\Span\{e_a, e_b\}$ on which each block is defined and $P_{\alpha \delta}$ is the permutation of the rows corresponding to axes $\alpha$ and $\delta$. As a convention for this work, every explicit matrix is given the standard basis ordering $(e_t, e_\mu, e_\nu, e_\xi)$. The parameters $g, b, h$ are placeholders for the supplied state's coordinates $\mu_0, \nu_0, \xi_0$, and we will call the vector $(g, b, h)$ the \textit{signature} of the protocol. The two blocks represent two types of coordinate dynamics. The first block, $M(1, g)_{t, \alpha}$, contains the probability of successful post-selection $p = \frac{1}{2}(t + g \alpha)$ and the update formula for coordinate $\alpha$ given by $\alpha \rightarrow (g + \alpha )/(1 + g \alpha)$ after normalization. The second block $M(b,h)_{(\beta, \gamma)}$ corresponds to a mixing of the coordinates $\beta$ and $\gamma$ with weights $b, h$. Due to the dynamical differences, we call the axis involved in the first block (here $e_\alpha$) the \textit{principal pumping axis} and the other two axes the \textit{secondary axes}. As a shorthand notation for the principal axis, we introduce $\delta \leftarrow g \alpha$. Consequently, the notation $G = G_{\delta \leftarrow g \alpha}(g, b, h)$ conveys all structurally relevant properties of a transfer matrix, but we may omit the full specification whenever it is not necessary and only write $G$.

The permutation operation $P_{\alpha \delta}$ is a crucial part of the expression. If $\alpha = \delta$, $G$ has due to $P_{\alpha \delta} = I$ a proper block structure in some ordered basis $(e_t, e_\alpha, e_\beta, e_\gamma)$. By an investigation presented in Appendix~\ref{sec:pumping_uniaxial}, we show that a protocol with a block-structured transfer matrix cannot sustain asymptotic fidelity improvement for interior supply states; only for states on a single face of the state space tetrahedron $\mathscr{F}_g \subset \partial \mathscr{T}$, such protocols can generate $F_\infty > \tfrac{1}{2}$.

In summary, the analysis of distillation protocols has been reduced to the methodologically simpler framework of transfer matrices. We proceed by establishing a classification of the entanglement pumping protocols, denoted by $\mathscr{P}(U, \Lambda_1)$, and demonstrate how to find optimal encodings under restricted system control, using the star-connectivity subalgebra $\mathfrak{g}_0$, taking the star-connected controllability restrictions from Sec.~\ref{sec:system_and_control_pump} as a concrete representative restriction.

\subsection{Classification of pumping protocols}
\label{sec:classification_of_pumping_protocols}
The transfer matrix $G$ of an entanglement-pumping scheme $\mathscr{P}(U,\Lambda_1)$ governs the entire distillation dynamics. In this section, we provide a complete classification of these transfer matrices, thereby fully categorizing all possible entanglement-pumping protocols. Rather than presenting the general argument first, we work through representative examples that show the defining features of each class. The systematic derivation of the classification summarized in Tab.~\ref{tab:classification_table_pumping_protocols} is given in Appendix~\ref{sec:classification_transfer_matrices}. It is based on the symplectic symmetries of the binary-vector representation of Pauli strings \cite{calderbank1998, hostens2005, dehaene2003, jansen2022, nielsen2010}.

In the following, we will consider mirrored gates $U$, which are generated on both partitions by a sequence of two-qubit Pauli strings $P_k \in \mathfrak{g}$ that belong to some dynamical Lie algebra $\mathfrak{g}$ and we write
\begin{align}
    U_A = \prod_{j=1}^k e^{-i \tfrac{\pi}{4} P_{j}}.
\end{align}
Moreover, we will use the shorthand notation $U = [P_1, \ldots, P_N]$ to denote pumping gates. Naturally, we require $k$ to be as small as possible; longer gates would make the system more susceptible to relaxation processes. As we will see, it is enough to focus on the treatment of the restricted DLA $\mathfrak{g} = \mathfrak{g}_0$, because we find plenty of suitable pumping protocols for $k \leq 2$. If required, one can easily repeat the procedure for other gate sets.

The evolution equation of the stabilizer coordinates and the defining equation of the transfer map in Eq.~\eqref{eq:update_rule_linear} already suggest that the transformation of the readout string $\Lambda_1$ under $U$ is an important determinant of the dynamic behavior of the protocol. To be more precise, the support of $\widetilde{\Lambda}_1$, i.e., the set of qubits on which the Pauli string acts with non-trivial factors, is the crucial element that governs the pumping dynamics. 

The first example is the protocol $\mathscr{P}_1 = \mathscr{P}([ZX, ZY], ZZ)$, where both two-qubit gates are the generators of the Lie algebra $\mathfrak{g}_0$. By simple inspection, we find that $[U, (ZZ)_1] = 0$, so $\widetilde{\Lambda}_1 = \Lambda_1 = (ZZ)_1$. Because $\widetilde{\Lambda}_1$ is only supported on the first pair, the strings $\widetilde{A}_2$ and $\widetilde{A}_2 \widetilde{\Lambda}_1$ will have the same substring on the second pair. Thus, for two stabilizer coordinates $\alpha, \beta$ with $\alpha = \langle A_2 \rangle$ and $\beta = \langle \widetilde{A}_2 \rangle$, the update by Eq.~\eqref{eq:update_rule_linear} has the form $\alpha' = \tfrac{1}{2}(1 + c_0) \beta$, where $c_0 = \langle \Lambda_1 \rangle$ is a constant given by a supply coordinate. In the case of the example $\mathscr{P}_1 = \mathscr{P}([ZX, ZY], ZZ)$, $c_0 = \langle ZZ_1 \rangle = \nu_0$, so the transfer matrix reads
\begin{align}
G_{\mathscr{P}_1} = \frac{1}{2}
\begin{pmatrix}
1 + \nu_0 & 0 & 0 & 0 \\
0 & 0 & 0 & 1 + \nu_0 \\
0 & 1 + \nu_0 & 0 & 0 \\
0 & 0 & 1 + \nu_0 & 0
\end{pmatrix}.
\end{align}
Hence, the protocol with this behavior only permutes the stabilizer coordinates with acceptance rate $\tfrac{1}{2}(1 + c_0)$, and we call protocols of this form \textit{permutation protocols}, which we will define as the first class of protocols in our classification.

A member of the second class is the pumping scheme $\mathscr{P}([XX, YY], ZZ)$. Here, the conjugated readout string $\widetilde{\Lambda}_1 = (ZZ)_2$ has only a support on the second pair and thus both terms in the evolution equation have the same substring on the first pair. Using an analogous argument as before, we find that $v_n = P_\pi v_0$ where $P_\pi$ is a permutation matrix that belongs to an entry permutation $\pi$ of the supply coordinate vector $v_0$. Consequently, protocols that have a $\widetilde{\Lambda}_1$ acting trivially on the first pair just discard the existing pair and replace it with a state similar to the supply pair $v_0$ -- appropriately, we will call this class of protocols the \textit{replacement protocols} (Class II).
\begingroup
\setlength{\tabcolsep}{9pt} 
\renewcommand{\arraystretch}{2} 

\setcellgapes{2pt}
\makegapedcells
\begin{table*}
    \centering
    \begin{tabular}{cccc}
        \toprule
        {Class} & {Description} & \makecell[l]{Transfer matrix / Output state} & {Condition(s)} \\
        \midrule
        I & \makecell{\textbf{Permutation} \\ \parbox[c]{4.5cm}{Pair coordinates are permuted (invariant fidelity)}} & $v' = P_\pi v$ & 
        $\restr{\smash{\widetilde{\Lambda}_1}}{2} = I $\\
        
        II & \makecell{\textbf{Replacement} \\ \parbox[c]{4.5cm}{Coordinates are replaced by a permutation of the supply pair coordinates ($F' = F_0$)}} & $v' = P_\pi v_0$ &
        $\restr{\smash{\widetilde{\Lambda}_1}}{1} = I $ \\
        
        III & \makecell{\textbf{Uniaxial pumping} \\ \parbox[c]{4.5cm}{Pumping protocol of the form $\alpha \leftarrow g \alpha$ (input and output axes coincide)}} & $G = \frac{1}{2} \left[ M(1, g)_{(t, \alpha)} \oplus M(b,h)_{(\beta, \gamma)} \right]$ & $\begin{aligned} 
        \restr{\smash{\widetilde{\Lambda}}_1}{1}, \restr{\smash{\widetilde{\Lambda}}_1}{2}  &\neq I \\ \left[\smash{\widetilde{\Lambda}_1}, U^\dagger \smash{\left( \restr{\smash{\widetilde{\Lambda}_1}}{2} \right)} U \right]_{a_2} &= 0 \end{aligned}$ \\
        
        IV & \makecell{\textbf{Biaxial pumping} \\ \parbox[c]{4.5cm}{Proper pumping protocol for any past-threshold supply. Input and output axes differ.}} & $G = \frac{1}{2} P_{\alpha \delta} \left[ M(1, g)_{(t, \alpha)} \oplus M(b,h)_{(\beta, \gamma)} \right]$ &$\begin{aligned} 
        \restr{\smash{\widetilde{\Lambda}}_1}{1}, \restr{\smash{\widetilde{\Lambda}}_1}{2}  &\neq I \\ \left[\smash{\widetilde{\Lambda}_1}, U^\dagger \smash{\left( \restr{\smash{\widetilde{\Lambda}_1}}{2} \right)} U \right]_{a_2} &\neq 0 \end{aligned}$ \\
        \bottomrule
    \end{tabular}
    \caption{Classification table for pumping protocols with Clifford group encoding $(U, \Lambda)$. There are four classes of protocols, only one of which provides entanglement pumping for any state fulfilling the distillation threshold $F_0 > \tfrac{1}{2}$ (class IV). The matrix $P_\pi$ denotes a permutation matrix corresponding to a row permutation $\pi: \mathbb{N}^n \rightarrow \mathbb{N}^n$. The notation $[\, \cdot \, , \, \cdot \, ]_{a_2}$ expresses the restriction of the commutator to qubit $a_2$.}
    \label{tab:classification_table_pumping_protocols}
\end{table*}
\endgroup

Let us now consider protocols whose $\widetilde{\Lambda}_1$ has support across both pairs. As examples, use $\mathscr{P}_3 = \mathscr{P}([XY], -YY)$ with $\widetilde{\Lambda}_{1} = (ZZ)_1(-YY)_2$ and the well-known Dür-protocol $\mathscr{P}_4 = \mathscr{P}([XI, IX, \cnot_{2 \rightarrow 1}], ZZ)$ according to Ref.~\cite{dur1999} with $\widetilde{\Lambda}_1 = (-YY)_1(-YY)_2$. We show in Appendix~\ref{sec:classification_transfer_matrices} that any protocol with $\tilde \Lambda_1$ with full support has a principal axis with the dynamical equation
\begin{align}
    \delta = \frac{1}{2} (g t + \alpha) ,
\end{align}
consistent with the example we studied beforehand. Here $\alpha$ is an input coordinate, $\delta$ an updated coordinate and $g$ is a coefficient in the transfer matrix. The structure of the principal axis is a major determinant for pumping capabilities, because we showed in Appendix~\ref{sec:pumping_uniaxial} that transfer matrices where $\alpha'$ and $\alpha$ belong to the same stabilizer coordinate (i.e.~the same axis $e_{\delta} = e_{\alpha}$) and $G$ attains a block structure, entanglement distillation capabilities are only provided for a limited subset of input states. This condition can be translated to the property
\begin{align}
    e_{\delta} = e_{\alpha} \quad \Longleftrightarrow \quad [\widetilde{\Lambda}_1, U^\dagger (\restr{\smash{\widetilde{\Lambda}_1}}{2}) U]_{a_2} = 0,
\end{align}
where $[\; \cdot \;, \; \cdot \;]_{a_2}$ is the commutator restricted to qubit $a_2$. Note that due to the mirror symmetry, one can also equivalently use $b_2$ instead of $a_2$. The arguments in the commutator determine the input and output axes, and therefore commuting elements imply identical axes. For the exemplary protocols $\mathscr{P}_3$ and $\mathscr{P}_4$ we get $U^\dagger \smash{\left( \restr{\smash{\widetilde{\Lambda}_1}}{2} \right)} U = (-YY)_2$ for $\mathscr{P}_3$ and $U^\dagger \smash{\left( \restr{\smash{\widetilde{\Lambda}_1}}{2} \right)} U = (XX)(ZZ)_2$ for $\mathscr{P}_4$, so they show uniaxial and biaxial pumping, respectively.

The classification including the membership conditions is summarized in Tab.~\ref{tab:classification_table_pumping_protocols}. Class IV is the only subset of distillation schemes that achieves $F_\infty > F_0$ for any supply state with sufficiently high initial fidelity and $g < 1$, so we call it the class of \textit{proper} pumping protocols. In Appendix~\ref{sec:pumping_conditions}, we find that $F_\infty > F_0$ holds for every class IV protocol with a supply state above the threshold $F_0 > \varphi^{-1}$ and $g < 1$, where $\varphi^{-1} \approx 0.618$ is the inverse golden ratio. This bound extends to $F_0 > \frac{1}{2}$ for optimal protocols discussed in Sec.~\ref{sec:optimal_protocols}. However, we will observe physical processes in Section~\ref{sec:entanglement_generation} that generate entangled pairs that are distillable with class III protocols. Since there are protocols in this class with $U$ consisting only of a single $(k=1)$ two-qubit gate in $\mathfrak{g}_0$, class III protocols offer faster distillation steps and are advantageous in this case.

\subsection{Optimal protocols}
\label{sec:optimal_protocols}
\begingroup
\setlength{\tabcolsep}{6pt} 
\renewcommand{\arraystretch}{1.5} 

\setcellgapes{2pt}
\makegapedcells
\begin{table*}
    \centering
    \begin{tabular}{cccc}
        \toprule
        {Signature $(g,b,h)$} & {Representative protocol} & {Pumping structure} & {Subclass}\\
        \midrule
        $(\mu_0, \nu_0, \xi_0)$ & {$\mathscr{P}([YY, YX], ZZ)$ }  & $\mu \leftarrow \mu_0 \xi$ & IV.5 \\
        $(\mu_0, \xi_0, \nu_0)$ & {$\mathscr{P}([ZX, ZY], -YY)$}  & $\xi \leftarrow \mu_0 \mu$ & IV.12 \\
        $(\nu_0, \mu_0, \xi_0)$ & {$\mathscr{P}([XX, YY], XX)$ }  & $\nu \leftarrow \nu_0 \xi$ & IV.16 \\
        $(\nu_0, \xi_0, \mu_0)$ & {$\mathscr{P}([XX, XY], -YY)$ }  & $\xi \leftarrow \nu_0 \mu$ & IV.10 \\
        $(\xi_0, \mu_0, \nu_0)$ &  {$\mathscr{P}([ZY, ZX], XX)$ }  & $\mu \leftarrow \xi_0 \xi$ & IV.1\\
        $(\xi_0, \nu_0, \mu_0)$ & {$\mathscr{P}([ZX, XY], ZZ)$ }  & $\nu \leftarrow \xi_0 \xi$ & IV.13 \\
        \midrule
        $(\xi_0, \mu_0, \nu_0)$ & {Dür et al.~\cite{dur1999}}  & $\nu \leftarrow \xi_0 \xi$ & IV.15\\
        \bottomrule
    \end{tabular}
    \caption{Examples of protocols for every possible signature $(g,b,h)$. The subclass is taken from the full classification of $\mathfrak{g}_0$ encodings in Tab.~\ref{tab:subclass-IV} in the appendix. For reference, the signature and structure of the Innsbruck protocol in Ref.~\cite{dur1999} are shown and identified as a IV.13 member.}
    \label{tab:pumping_signature_table}
\end{table*}
\endgroup
An entanglement pumping protocol is optimal when it produces the highest asymptotic fidelity $F_\infty$ with minimal protocol iterations. There are two degrees of freedom for transfer matrices, namely the pumping structure $\delta \leftarrow g \alpha$ and the signature. We will begin by showing that the pumping structure $\delta \leftarrow g \alpha$ is not relevant in terms of the asymptotic fidelity $F_\infty$. To show this, we take two transfer matrices $G_1$ with a structure $\delta \leftarrow g \alpha$. Let a second transfer matrix be related to $G_1$ by a permutation matrix $P \equiv P_\pi$ so that $G_2 = P G_1 P^T$ has a pumping structure $\pi(\delta) \leftarrow g \pi(\alpha)$ altered by a permutation $\pi(\cdot)$. For an eigenvector $v$ of $G_1$ with eigenvalue $\lambda$ follows that
\begin{align}
    G_2(P v) = P G_1 P^T P v = \lambda (Pv),
\end{align}
so $G_1$ and $G_2$ have the same eigenvectors up to permutations. Since the fidelity metric is permutationally invariant $F(v) = F(P v)$, the pumping structure of a protocol is not a determining factor of $F_\infty$.

Instead, the optimal $F_\infty$ is determined only by the signature of $G$, $(g,b,h)$. By observation in numerical investigation, we conjecture that $F_\infty$ is maximized if
\begin{align}
\label{eq:min_fuel_conjecture}
    g \leq b \leq h.
\end{align}
A rigorous, analytical proof of this claim is challenging as it requires determining the maximum sum of the components of the leading eigenvector of a matrix with a given structure over all possible combinations of its elements. However, to motivate the statement that $g$ should be minimal, we can compute a lower bound on the output infidelity (see Appendix~\ref{sec:bound_F_infty}) and obtain
\begin{align}
    1 - F_\infty \geq &\frac{1 + g - b - h}{2(1 + g)} \nonumber \\&= \frac{1 +2 g - \Sigma_0}{2(1 + g)} \eqdef \varepsilon_\text{min}(g),
\end{align}
where $\Sigma_0 = g + h + b \equiv \text{const} > 0$ is a constant scalar describing the quality of the supply pair. The derivative
\begin{align}
    \dv{\varepsilon_\text{min}}{g}(g) = \frac{1 +\Sigma_0}{2(1 + g)^2} \geq 0,
\end{align}
is positive across all $g$ and therefore the bound $\varepsilon_\text{min}(g)$ is a strictly increasing function with increasing $g$.

\begin{figure*}
    \centering
    \includegraphics[width=0.65\linewidth]{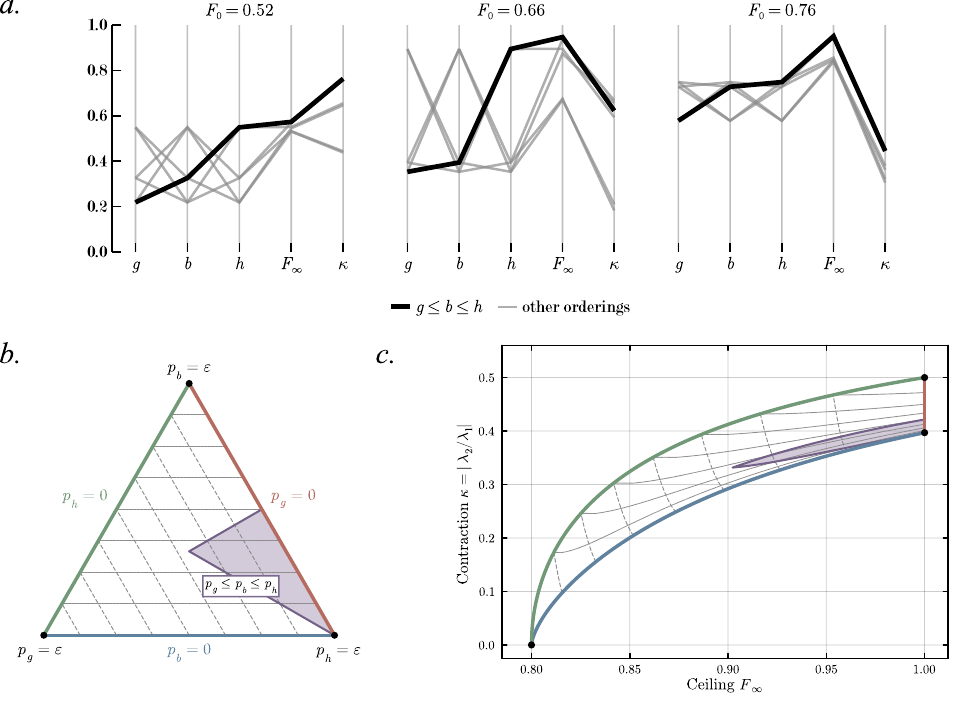}
    \caption{Optimal signature of a pumping scheme. (a) Parallel coordinates plot (PCP) of the performance tuple $(F_\infty, \kappa)$ with respect to three different supply states $v_i^0$. For every supply state, all possible signatures $(g, b, h)$ are displayed, and the signature fulfilling $g \leq b \leq h$ is highlighted. One observes that the highlighted signature always belongs to the highest $F_\infty$, but not necessarily to the best contraction value $\kappa$. (b) The 2-simplex of input states with equal fidelity $\Delta_{F_0}$ (see Fig.~\ref{fig:tetrahedron}) and its mapping into the $(F_\infty, \kappa)$ diagram. At the vertices, only two populations are non-zero, where the target population is $p_0 = F_0$, and the other one corresponds to the infidelity $\varepsilon \defeq 1 - F_0$. The blue boundary constitutes the Pareto front of the optimization problem. States satisfying the conjecture $g \leq b \leq h$ (i.e., $p_g \leq p_b \leq p_h$) are indicated by the shaded region.}
    \label{fig:optimal_protocol} 
\end{figure*}

Figure~\ref{fig:optimal_protocol} summarizes how the choice of signature determines the performance tuple
$(F_\infty,\kappa)$, where the contraction parameter $\kappa \defeq \abs{\lambda_2 / \lambda_\star}$ describes the inverse convergence rate towards the asymptotic fidelity. In Fig.~\ref{fig:optimal_protocol}(a), all six signatures are evaluated for three supply states of increasing fidelity. In each case, the signature satisfying $g\le b\le h$ reaches the highest $F_\infty$, supporting conjecture \eqref{eq:min_fuel_conjecture}. However, it generally does not produce the smallest contraction $\kappa$, as demonstrated by the three examples in Fig.~\ref{fig:optimal_protocol}(a). Thus, optimizing for maximum fidelity $F_\infty$ and convergence rate $\kappa$ are in general competing objectives. In Fig.~\ref{fig:optimal_protocol}(b), we resolve this trade-off over an iso-fidelity simplex $\Delta_{F_0}$, the subspace of states with fidelity $F \equiv F_0$ (illustrated in the state space tetrahedron in Fig.~\ref{fig:tetrahedron}). For every stabilizer coordinate of $\rho_0$, $c_x, c_y, c_z \in \{\mu_0, \nu_0, \xi_0\}$ we find a Bell population $p_{c_x}$ with
\begin{align}
    p_x = \frac{1}{4} \left( 1 + c_x - c_y - c_z \right),
\end{align}
given by Eq.~\eqref{eq:populations_expressed_in_coords}. The inverse transformation is $c_x = (p_0 + p_x) - (p_y + p_z) = 2 (p_0 + p_x) - 1$, where the last equality is justified by $\sum_i p_{c_i} = 1$. This implies that differences remain invariant up to a scaling factor $c_x - c_y = 2(p_x - p_y)$, allowing us to conclude that our conjecture in Eq.~\eqref{eq:min_fuel_conjecture} is equivalent to
\begin{align}
    p_g \leq p_b \leq p_h.
\end{align}
The edges of $\Delta_{F_0}$ correspond to supply states $\rho_0$ with three non-zero populations and the vertices correspond to states with two non-zero populations; an analog to the state shape of Ref.~\cite{dur1999}. How $\Delta_{F_0}$ is oriented within the tetrahedron, i.e.~which edge corresponds to $p_g=0, p_b=0, p_h=0$, depends on the protocol signature $(g,b,h)$ and should be chosen such that the supply state coordinates are as close to the upper vertex $p_h = \varepsilon$ as possible. This selection rule is a geometric perspective on the problem equivalent to condition \eqref{eq:min_fuel_conjecture}. To conclude, a protocol is optimal with respect to a supply pair with state $\rho_0$ if the protocol signature satisfies the conjectured ordering $g \leq b \leq h$. For our dynamical algebra $\mathfrak{g}_0$, all possible signatures can be constructed (see Tab.~\ref{tab:pumping_signature_table}), but this might not be the case for arbitrary control restrictions.

\subsection{Composite protocols}
In the previous sections, we analyzed the pumping behavior of a given protocol and classified it by its transfer matrix. We noted that, except for special cases, only class IV protocols are useful, and almost all of them require two or more two-qubit gates from the gate set defined by our DLA to establish the encoding.

Another way of creating class IV pumping schemes is the \textit{composition} of protocols. Here, we change the encoding after every pumping iteration (see Fig.~\ref{fig:entanglement_protocols}(b)); therefore, the effective transfer matrix after a cycle of an alternation between $k$ protocols is given by their product
\begin{align}
    G = G_k G_{k-1} \cdots G_2G_1.
\end{align}
We show an example that combining faulty elements yields a protocol with improved performance. Take two transfer matrices $G_1$ and $G_2$ of class III. Due to their block structure, they fail to provide a pumping dynamics for interior supply states $\mathscr{T}^\circ$ and therefore for the vast majority of states. Assume that those two matrices are related by a transposition of coordinates $P$, i.e.~$P G_1 P^T = P G_1 P = G_2$. Then the effective transfer matrix is
\begin{align}
    G = G_2 G_1 = (P G_1 P) G_1 = (P G_1)^2\; \in \;  \text{Class IV}.
\end{align}
So, after two steps of the composed protocol, we have an effective class IV dynamics as given by Eq.~\eqref{eq:transfer_matrix_structure} with $P \neq I$. In the case of our assumed dynamical Lie algebra $\mathfrak{g}_0$, we find several encodings in class III that are generated by only one two-qubit gate (see Tab.~\ref{tab:subclass-III}). In cases where the execution time of a single two-qubit gate is significant compared to the overhead of the other elements of the entanglement pumping algorithm (generation time of the supply state, measurement, post-processing, classical overhead, etc.), the composition technique might perform better than one-cycle protocols.

\begin{figure}
    \centering
    \includegraphics[width=0.95\linewidth]{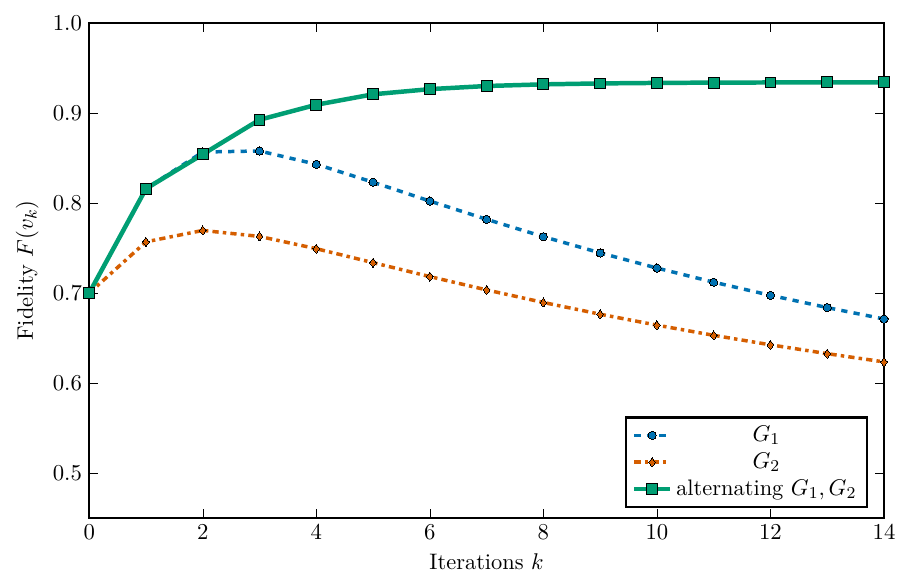}
    \caption{Demonstration of the distillation properties of a composed protocol represented by $G$ relative to its components $G_1, G_2$ for a supply state $v_0 = (1, 0.45, 0.675, 0.675)^T \in \mathscr{T}^\circ$ and $F_0 = 0.7$.}
    \label{fig:composed_protocol_demo}
\end{figure}

As an example, we examine the two protocols $\mathscr{P}_1 = \mathscr{P}([YX], ZZ)$, $\mathscr{P}_1 = \mathscr{P}([YY], ZZ)$. Individually, they are class III protocols, but with the transposition $P \equiv P_{(\mu \xi)}$, their transfer matrices are related by
\begin{align}
    G_1 &= 
    \frac{1}{2}\begin{pmatrix}
        1 & \mu_0 & 0 & 0\\
        \mu_0 & 1 & 0 & 0\\
        0 & 0 & \nu_0 & \xi_0\\
        0 & 0 & \xi_0 & \nu_0
    \end{pmatrix}, \;
    G_2 = 
    \frac{1}{2}\begin{pmatrix}
        1 & 0 & 0 & \mu_0\\
        0 & \nu_0 & \xi_0 & 0\\
        0 & \xi_0 & \nu_0 & 0\\
        \mu_0 & 0 & 0 & 1
    \end{pmatrix}, \nonumber \\
    G &\defeq G_2 G_1 = [P G_1]^2 =\frac{1}{4}
    \left[\begin{pmatrix}
        1 & \mu_0 & 0 & 0\\
        0 & 0 & \xi_0 & \nu_0\\
        0 & 0 & \nu_0 & \xi_0\\
        \mu_0 & 1 & 0 & 0
    \end{pmatrix}\right]^2.
\end{align}
In Fig.~\ref{fig:composed_protocol_demo} we computed the entanglement pumping dynamics generated by $G_1$, $G_2$ and $G$ for an interior supply state $v_0 \in \mathscr{T}^\circ$. Here we can clearly observe that the composed protocol is able to carry out entanglement distillation with only one two-qubit gate as an encoding, whereas its components converge towards fidelity $\tfrac{1}{2}$. Therefore, the composition of protocols is a method to create functioning pumping schemes, which increases the number of options for a protocol design. 

In summary, the asymptotic fidelity of a biaxial protocol is determined by its signature and is maximized for $g\le b\le h$ as found numerically. For a given register, one therefore only needs to determine the accessible gates, classify the resulting transfer matrices, and select a biaxial protocol with the ordered signature. For $\mathfrak g_0$, all six signatures are realized with at most two Clifford gates (Tab.~\ref{tab:pumping_signature_table}), so the restricted control we considered lowers no asymptotic fidelity. Since this star-shaped structure is common to many solid-state spin registers, we expect this conclusion to hold broadly. For other control restrictions, missing signatures may still be obtained by composite protocols.

\section{Implementation in Entanglement Generation}
\label{sec:entanglement_generation}
In this section, we present two heralded entanglement-generation schemes for generating entangled pairs and discuss scenarios in which entanglement distillation can be beneficial. Uncontrolled experimental fluctuations and coupling to environmental degrees of freedom can introduce uncertainty into the heralded preparation, resulting in an experimental state $\rho_{\mathrm{exp}}$ that is mixed and deviates from the target state. Specifically, in photon-mediated schemes, photonic degrees of freedom are first entangled with two separated qubits, after which the photons are transmitted and interfered on the balanced beam splitter and are finally detected in distinct output modes. The selected detection patterns then herald an entangled pair~\cite{Reiserer2015, Beukers2024}. Imperfections arising during photon generation, transmission, interference, and detection can reduce the fidelity of the resulting heralded state. In the following, we introduce the relevant imperfections, briefly discuss their effects, and assess the potential benefit of entanglement distillation.

\subsection{Phase-encoding based entangling proposal}
In phase-encoding schemes, a single photon is emitted from the source and coherently split into two paths, denoted by $A$ and $B$. The photon in each path then interacts with one of two remote qubits, which are initialized in $(\ket{0}+\ket{1})/\sqrt{2}$, thereby imprinting a spin-dependent $\pi$ phase shift on the corresponding optical path. The two paths are subsequently recombined on a beam splitter, and detection at either output port heralds a Bell state of the remote qubits, namely $\ket{\psi}_A=\ket{\Psi^-}$ and $\ket{\psi}_B=\ket{\Phi^-}$~\cite{Duan2004, Hu2008, omlor2025}. The two heralded states can then be aligned by a local operation.
The ideal derivation assumes a stable relative optical phase and indistinguishable photonic modes. In practice, phase fluctuations and imperfect mode overlap reduce the coherence of the heralded spin state. Averaging over the phase fluctuations and taking into account the imperfect overlap, the resulting mixed state can be written as
\begin{align}
    \rho_0^{\mathrm{(phase)}} &=
    \frac{1+\operatorname{Re}\mathcal{V}}{2}\dyad{\Psi^-}{\Psi^-}
    +\frac{1-\operatorname{Re}\mathcal{V}}{2}\dyad{\Phi^-}{\Phi^-} \nonumber \\
    &\phantom{=} +\frac{i\operatorname{Im}\mathcal{V}}{2}\dyad{\Psi^-}{\Phi^-}
    -\frac{i\operatorname{Im}\mathcal{V}}{2}\dyad{\Phi^-}{\Psi^-}
\end{align}
with $\mathcal{V}=\mu_{\text{ov}} e^{i\delta_0}e^{-\sigma_{\phi}^2/2}$ is a factor induced by overlap $\mu$ and phase difference $\delta_\phi$.  

First, we model the relative phase as $\delta_\phi(t)=\delta_0+\delta(t)$, where $\delta_0$ is a quasi-static offset and $\delta(t)$ denotes zero-mean fluctuations, which are assumed to follow a Gaussian distribution $ \delta_\phi\sim\mathcal{N}(\delta_0,\sigma_\phi^2)$. The offset mainly originates from static path-length mismatch and slow thermal or strain drifts. $\delta(t)$ accounts for residual phase fluctuations that occur during or across the attempts, which arise from mechanical vibrations, laser noise, and other environmental perturbations. Time-bin encoding can further suppress phase errors by using closely spaced early and late photons as resources, such that a short interval between two bins ensures better stability of the relative phase~\cite{Hensen2015, Stas2022, Knaut2024}.
Moreover, the active stabilization suppresses both slow drifts and dynamical phase fluctuations within the feedback bandwidth, leaving a finite residual phase noise~\cite{Stolk2025, Nardelli2025}. To this end, a co-propagating classical reference field is used to measure the differential optical phase, and the following correction may be implemented either optically or through an equivalent local phase rotation on one spin. The residual phase is now replaced by a new distribution $\tilde{\delta}_\phi(t)$ with smaller offset $\tilde{\delta}_0$ and uncertainty $\tilde{\sigma}_\phi$. 

Second, the complex mode overlap $\mu_{\text{ov}}$ quantifies the coherence between the photonic modes associated with the two remote arms. The overlap can be reduced by mismatches in the spectral, temporal, spatial, or polarization degrees of freedom. In reflection-based schemes, differences between the two registers can lead to distinct spin-conditioned reflection responses and hence different output photon wave packets. In addition, intrinsic pure dephasing of the optically excited state can reduce photon indistinguishability by degrading the optical coherence of the scattered field~\cite{Grange2015}. 
The deterministic phase can be compensated by an appropriate local spin rotation or an equivalent optical phase correction, making the effective overlap real and thereby maximizing the fidelity for a fixed $|\mu|$. The phase can be calibrated using an interferometric measurement or inferred from prior spin-state tomography.

Although deterministic phase offsets and the complex mode overlap can, in principle, be compensated through local rotations or active optical phase stabilization, residual phase noise with $\tilde{\sigma}_\phi>0$ and incomplete photonic indistinguishability with $|\mu_{\text{ov}}|<1$ impose intrinsic limitations on the attainable fidelity $F=(1+|\tilde{\mathcal{V}}|)/2$ with $\tilde{\mathcal{V}}=\mu_{\text{ov}} e^{i\tilde{\delta}_0}e^{-\tilde{\sigma}_{\phi}^2/2}$. Therefore, entanglement distillation provides a systematic means of suppressing the remaining Bell-state errors and producing entangled pairs of higher fidelity. Such processing is particularly relevant to long-distance entanglement distribution, for which maintaining phase stability and mode matching across spatially separated nodes becomes increasingly demanding. 

Under ideal phase compensation, we first apply a local Clifford operation $I\otimes Y$ to map the target state $|\Psi^-\rangle$ to $|\Phi^+\rangle$, consistent with the stabilizer convention $M=XX$ and $N=ZZ$. The experimental state then becomes $F|\Phi^+\rangle\langle\Phi^+|+(1-F)|\Psi^+\rangle\langle\Psi^+|$, giving $(\mu_0,\nu_0,\xi_0)=(1,2F-1,2F-1)$. All three coordinates are positive for $F>1/2$. To achieve the condition $g\leq b\leq h$, one candidate local unitary $U_A=\exp\left(-i\pi ZX/4\right)\exp\left(-i\pi XY/4\right)$ gives the signature assignment $(g,b,h)=(\xi_0,\nu_0,\mu_0)$, which is protocol $\mathscr{P}([ZX,XY],ZZ)$ of subclass IV.13 in Table~\ref{tab:subclass-IV}. After $n$ pumping rounds, the fidelity and cumulative success probability are $F_n=1/(1+R^{k_n})$ and $P_n=F^{n+1}(1+R^{k_n})$, respectively, where $R=(1-F)/F$ and $k_n=n+1-\left\lfloor(n+1)/3\right\rfloor$. For $F>1/2$, one has $R<1$ and hence $F_n\rightarrow1$ as $n\rightarrow\infty$, consistent with $p_g=0$ for this supply state, at the cost of a decreasing cumulative success probability.
Interestingly, the first pumping step also suppresses the residual coherent phase error. The retained pair after the first successful distillation round becomes
\begin{align}
    \rho_1^{\mathrm{(phase)}}=\frac{{(1+\operatorname{Re}\tilde{\mathcal{V}})^2\dyad{\Phi^+}{\Phi^+}+(1-\operatorname{Re}\tilde{\mathcal{V}})^2\dyad{\Psi^-}{\Psi^-}}}{2[1+(\operatorname{Re}\tilde{\mathcal{V}})^2]}.
\end{align}
The off-diagonal terms vanish after postselection, showing that the protocol suppresses the residual coherent phase noise.

\subsection{Emission-based entangling proposal}
In this section, we will introduce a different proposal that uses two emitters to prepare an entangled pair. At first, both spins $M=A, B$ are initialized as $\sqrt{1-P_e}\ket{0}_M+\sqrt{P_e}\ket{1}_M$, where $P_e\ll1$ and $\ket{1}_M$ is an optically bright state. State-selective photon emission implements the spin-photon encoding $\ket{1}_M\rightarrow\ket{1}_Ma_M^\dagger$,  whereas the dark state $\ket{0}_M$ does not couple to the optical mode~\cite{barrett2005}. An equivalent mapping can also be realized through a cavity-assisted Raman process in a three-level $\Lambda$ system, where the optical cavity couples the bright state to an excited state and the Raman transition generates a photon conditional on the spin occupying $\ket{1}_M$~\cite{bose1999, cabrillo1999}. Then the emitted photons are guided to interfere on a balanced beam splitter to erase the which-path information. Detecting one photon heralds that two remote spins are prepared in a Bell state $\ket{\Psi^+}$, up to a local $Z$ rotation, with success probability $2(1-P_e)P_e$. This protocol has been performed on various solid-state systems, such as quantum dots~\cite{delteil2016} and color centers~\cite{sun2018, pompili2021}. 

If the detector cannot resolve photon number, a double-emission event will be incorrectly accepted as a successful heralding event, thereby introducing an intrinsic $\ket{11}$ error component with probability $P_e^2$. This error can be substantially suppressed using photon-number-resolving detectors, although it cannot be completely eliminated at finite detection efficiency. In addition, the loss of one photon in a two-photon event can make such an event indistinguishable from a useful single-photon detection. The heralded state can therefore be written as $\left(P_1\dyad{\Psi^+}+P_2\dyad{11}\right)/P_{\mathrm{click}}$, where $P_{\mathrm{click}}=P_1+P_2$ is the total click probability. If only intrinsic double emission is considered, the two probabilities are $P_1=2P_e(1-P_e)$ and $P_2=P_e^2$.
Similarly, considering the relative phase fluctuation $\delta_\phi$ and non-unit overlap $\mu$, the Bell component $\dyad{\Psi^+}$ will be degraded due to errors into the state 
\begin{align}
    \rho'_{\mathrm{phase}} &=
    \frac{1+\operatorname{Re}\mathcal{V}}{2}\dyad{\Psi^+}{\Psi^+}
    +\frac{1-\operatorname{Re}\mathcal{V}}{2}\dyad{\Psi^-}{\Psi^-} \nonumber \\
    &\phantom{=} +\frac{i\operatorname{Im}\mathcal{V}}{2}\dyad{\Psi^+}{\Psi^-}
    -\frac{i\operatorname{Im}\mathcal{V}}{2}\dyad{\Psi^-}{\Psi^+}
\end{align}
giving the final state as
\begin{align}
    \rho_0^{(\mathrm{emission})} = \frac{1}{P_{\mathrm{click}}} \left( P_1 \rho'_{\mathrm{phase}} + P_2\dyad{11}\right).
\end{align}

Under ideal phase compensation and after the local operation $I\otimes X$, the heralded state reads $F\dyad{\Phi^+} + (1-F-\epsilon)\dyad{\Phi^-} + \epsilon\dyad{10}$, with $\epsilon=P_2/P_{\mathrm{click}}\ll 1$. 
With $M=XX$ and $N=ZZ$, its coordinates are $\mu_0=\xi_0=2F-1+\epsilon$ and $\nu_0=1-2\epsilon$.
Thus, using the protocol $\mathcal{P}([IZ,ZY],XX)$ of subclass IV.1 with the signature $(g,b,h)=(\xi_0,\mu_0,\nu_0)$, the transfer matrix is
\begin{align}
G(\epsilon)=\frac{1}{2}
    \begin{pmatrix}
    1 & 0 & 0 & \xi_0 \\
    \xi_0 & 0 & 0 & 1 \\
    0 & \nu_0 & \mu_0 & 0 \\
    0 & \mu_0 & \nu_0 & 0
    \end{pmatrix}.
\end{align}
Treating the two-photon events as a perturbation term, we can expand the Perron-Frobenius eigenvalue as $\lambda_\star=\lambda_0+\lambda_1\epsilon+\lambda_2\epsilon^2+O(\epsilon^3)$, in which $\lambda_0=F$ is the value when $\epsilon=0$. Solving the equation $\det(G-\lambda I)=0$, we have $\lambda_1=0$ and $\lambda_2=(1-F)^2/[4F(2F-1)]$. Then Perron-Frobenius eigenvector is $v_\star=\{1,\mu_\star=(\xi_0^2+2\lambda_\star-1)/(2\xi_0\lambda_\star), \nu_\star=\nu_0\mu_\star/(2\lambda_\star-\mu_0),\xi_\star=(2\lambda_\star-1)/\xi_0\}^T$, which gives the limit of the fidelity as
\begin{align}
    F_\infty = 1-\frac{F^2-F+1}{2F(2F-1)}\epsilon + \mathcal{O}(\epsilon^2).
\end{align}
Typically, two-photon events limit the achievable fidelity, and as $\epsilon\rightarrow0$, the fidelity approaches unity $F_\infty\rightarrow1$. Remarkably, distillation compresses the two-photon events admixture below its initial value if $F>0.768$, yielding $F_\infty>1-\epsilon$.

As discussed above, phase fluctuations, imperfect photonic-mode overlap, and double-emission events limit the fidelity of the entangled state, while entanglement distillation can be used to suppress these error contributions. In practical quantum networks, entanglement generation may also be affected by other imperfections, such as memory decoherence, imperfect photonic interference, and local-gate and measurement errors. The effectiveness of distillation in their presence will depend on the specific network architecture, entanglement-generation scheme, and experimental conditions, and therefore requires further investigation.
%
%
%
\section{Conclusion and outlook}
\label{sec:conclusion}
We have developed a general framework for analyzing entanglement pumping protocols and applied it to a spin network with restricted control properties. As a representative example, we consider NV center quantum registers with control capabilities provided only by the central electron spin. In a single register, one optically active electron spin is driven, initialized, and read out, while the purified pair is stored in hyperfine-coupled nuclear spins, providing a concrete setting with restricted local operations. Although the electron--nuclear register of a color center serves as the guiding example, the analysis relies only on the star-shaped connectivity and indirect electron-mediated control of the memory qubits. It therefore extends to other central-spin architectures with the same control properties.

To reveal the structure of entanglement pumping, we developed a stabilizer-coordinate framework, in which coordinates $(\mu,\nu,\xi)$ are defined by the expectation values of the three non-trivial stabilizers associated with the target Bell state. Together with a normalization coordinate, each pumping step could be expressed as an update of all coordinates via a linear transfer map $G$. This representation has three consequences. First, the asymptotic fidelity and the convergence rate under repeated pumping follow directly from the eigensystem of the transfer matrix. The performance of a protocol is therefore obtained from a four-dimensional eigenvalue problem rather than from iterated density-matrix simulations. Second, the coordinate update rule yields a criterion for identifying non-trivial pumping protocols. Third, the asymptotic fidelity depends explicitly on the assignment of the supply-state coordinates. 
This framework is therefore well suited to entanglement pumping with limited control. Once the accessible Clifford gates are determined from the system controllability, their pumping performance can be compared directly, and the best protocol can be selected.

Several aspects merit further investigation. First, we restricted the encodings to mirrored Clifford gates, since the stabilizer coordinates form a closed set of dynamical variables only for encodings that are mirrored between the partitions $A$ and $B$. Asymmetric encodings render the output dependent on the Bell-basis coherences of the supply state and thereby break this closure; they may therefore give rise to pumping dynamics outside the present classification. A corresponding framework for asymmetric two-qubit gates would require, instead of the three stabilizer coordinates, all 15 coordinates $c_P = \langle P \rangle$ given by the expectation values of the non-identity two-qubit Pauli strings $P$, which form a basis of $\mathfrak{su}(4)$. Alternatively, entanglement pumping schemes based on asymmetric or non-Clifford gates could be studied by parameterizing the Lie group $e^{\mathfrak{g}}$ of the control problem through a suitable decomposition, e.g.~the $KAK$ decomposition~\cite{wang2025, kokcu2022, khaneja2001}, and optimizing the encoding $U[\vec{\theta}]$ with respect to the performance objectives $(1-F, \kappa)$ over the free parameters $\vec{\theta}$. Such an approach will likely require additional symmetries or structure to be imposed on the encoding circuit, since the number of free parameters grows quickly: two independent two-qubit gates already involve $2[\dim(\mathrm{SU}(4))] = 30$ real parameters. Second, our identification of the signatures of optimal protocols rests solely on numerical evidence. In particular, we found numerically that the ordering $g \leq b \leq h$ yields the maximal asymptotic fidelity in the limit of infinitely many iterations. An analytical study of the eigensystem of biaxial transfer matrices could establish a sharper condition for optimality, including the practically relevant case of a finite number of iterations $N < \infty$. Third, an interesting avenue for future research is the systematic generalization of this classification scheme to multi-pair configurations involving general $n$-to-$k$ pair entanglement pumping protocols \cite{kim2026, jansen2022}. Finally, incorporating relaxation and dephasing into the formalism would be a valuable extension: in the presence of noise, the maximal fidelity is no longer attained asymptotically, and a finite optimal number of iterations can be identified. Ref.~\cite{addala2025} offers a feasible route that may be helpful here, in which noisy circuits can be simulated efficiently by tracking only the stabilizer signs of each Bell pair.

In summary, restricted control in centrally controlled registers is not an obstacle to entanglement pumping. The design problem reduces to identifying a biaxial transfer-matrix structure that is reachable with the native gates, or composing one from uniaxial building blocks, and to choosing the signature that is appropriate for the shape (Bell populations) of the supply state. This provides a systematic route to entanglement distillation on few-qubit nodes.

\acknowledgments{We thank Martin Korzeczek for helpful discussions. This work is supported by the BMBF under the funding program ‘quantum
technologies—from basic research to market’ in the project Spinning (project No. 13N16215) and project
CoGeQ (project No. 13N16101). This work was supported by EU-project C-QuENS (Grant No. 101135359)
and EU-Project SPINUS (Grant No. 101135699), by Ulm University via Science Sphere and the Ministry for Science, Research and the Arts Baden-W{\"u}rttemberg via the project "MolQuant" in the framework of the BEGIN Framework.
}

\bibliography{literature.bib}

%
%
%
%
%
%
%

\onecolumngrid
\appendix
\section{The stabilizer coordinates}
\label{sec:stabilizer_coordinates}
In this work, we analyze entanglement distillation protocols with mirrored encoding gates, and thus our description of the entanglement pumping process in coordinates defined in terms of matched strings becomes very efficient. In this section, we properly define the notion of matched strings and mirrored gates. Furthermore, we explicitly show why our introduced stabilizer coordinates define a closed coordinate system (Proposition~\ref{prop:matched_string_decompositon}). 
\begin{definition}[Mirror operator $\tau$]
For a Pauli string $P = \left( \bigotimes_{j=1}^n P_{a_j} \right) \otimes \left( \bigotimes_{j=1}^n P_{b_j} \right) \in \mathcal{P}_{2n}$ partitioned between two subsystems $A$ and $B$, the mirror operator $\tau$ permutes the tensor factors by swapping $a_j \leftrightarrow b_j$ for every $j$.
\end{definition}
For instance, the action of $\tau$ on a Pauli string $U_{a_1} V_{b_1} W_{a_2}$ gives $\tau(U_{a_1} V_{b_1} W_{a_2}) = V_{a_1} U_{b_1} W_{a_2}$. Also note that the definition of $\tau$ can be extended to gates of the form $U = \exp(-i \tfrac{\phi}{2} P)$ by
\begin{align}
    \tau(U) = \cos \left( \frac{\phi}{2} \right) I -i \sin \left( \frac{\phi}{2} \right) \tau(P).
\end{align}
\begin{definition}[Matched strings and symmetric gates]
A Pauli string $P$ is called a matched string if
\begin{align}
    \tau(P) \propto P.
\end{align}
We denote the set of matched strings by $\mathcal{M}$. The set of matched strings is closed under multiplication, and any two matched strings commute. A gate $U = U_A U_B$ is called symmetric, if
\begin{align}
    U_B = \tau(U_A)^*.
\end{align}
\end{definition}
\begin{proposition}
Symmetric gates within the Clifford group $\varphi_U(P): \mathcal{P} \rightarrow \mathcal{P}, P \mapsto U P U^\dagger$ map matched strings to matched strings.
\end{proposition}
\begin{proof}
It is enough to show this for a matched substring on a single pair $P_{a_j} P_{b_j}$. The gate is symmetric so $U = U_A U_B = U_A \tau(U_A)^*$. Assume that $U_A P_{a_j} U_A^\dagger = Q_{a_j}$. For any Pauli string, the conjugation only affects the sign $P^* = c_P P$ with $c_P = \pm1$. Thus,
\begin{align}
    U_B P_{b_j} U_B^\dagger &= \tau(U_A)^* \tau(P_{a_j}) \tau(U_A^\dagger)^* = \tau(U_A^* P_{a_j} U_A^T) \nonumber \\
    &= c_P\tau(U_A P_{a_j} U_A^\dagger)^* = c_P \tau(Q_{a_j})^* \nonumber \\
    &= c_P c_Q Q_{b_j}.
\end{align}
We determined that $\varphi_U(P_{a_j}P_{b_j}) = \pm Q_{a_j} Q_{b_j}$. Since this holds for any pair $j$, we can conclude that $\varphi_U: \mathcal{M} \rightarrow \mathcal{M}$.
\end{proof}
\begin{proposition}
\label{prop:matched_string_decompositon}
Let $T$ be a Hermitian matched string with an overall sign $s(T) \in \{\pm 1\}$ and let $\rho = \bigotimes_{j=1}^n \rho_j$ be a product state of $n$ two-qubit pairs. The expectation value of $T$ is given by
\begin{align}
    \langle T \rangle_\rho = s(T) \prod_j \langle \restr{T}{j} \rangle_{\rho_j},
\end{align}
Here, $\restr{T}{j}$ is the substring of $T$ that is supported on pair $j$. Eventually, the expectation value can be expressed as a product of the matched correlators because
\begin{align}
    \label{eq:expectation_value_components}
    \langle \restr{T}{j} \rangle_{\rho_j} = \langle Q_{a_j} Q_{b_j} \rangle_{\rho_j} =
    \begin{cases}
        \phantom{-}1, &Q = I,\\
        \phantom{-}\mu_j, &Q = X,\\
        \phantom{-}\nu_j, &Q = Z,\\
        -\xi_j, &Q = Y.
    \end{cases}
\end{align}
\end{proposition}
\begin{proof}
Any Pauli string trivially factorizes across the pairs $j$ into $T = s(T) \prod_j \restr{T}{j}$. Since $\rho$ is a product state across the two-qubit pairs, the expectation value evaluates to
\begin{align}
    \langle T \rangle_\rho = \Tr \left\{ T \rho \right\} = s(T) \Tr \bigg\{ \bigotimes_j \restr{T}{j} \rho_j \bigg\} = s(T) \prod_j \Tr{ \restr{T}{j} \rho_j} = s(T) \prod_j \langle \restr{T}{j} \rangle_{\rho_j}.
\end{align}
For a matched string $T$, $T = T_A \otimes T_B \propto T_A \otimes \tau(T_A)$, its restriction onto a single pair is (up to a sign) $\restr{T}{j} \propto Q_{a_j} Q_{b_j}$ for a Pauli matrix $Q$. Since every substring is a matched string acting on a single pair, the expectation value can be written in terms of a product of single-pair coordinates $(\mu_j, \nu_j, \xi_j)$ with the rule according to Eq.~\eqref{eq:expectation_value_components}.
\end{proof}
With respect to the dynamics of the distillation protocol shown in the main text (especially Eq.~\ref{eq:stabilizer_update_rule}) and the properties of matched strings and symmetric gates in this section, this analysis suggests that the coordinates $(\mu, \nu, \xi)$ of an entangled pair constitute suitable coordinates for the description of distillation protocols with symmetric operations on the partitions $A$ and $B$, because the distilled entangled pair(s) have coordinates $(\mu', \nu', \xi')$ that only depend on the stabilizer coordinates of the input pairs.

Symmetric gate setups are a major subclass of distillation protocols. For asymmetric distillation protocols, this description fails, because the output fidelity is in general dependent on off-diagonal elements of $\rho$ in the Bell basis and therefore also depends on the expectation value of unmatched strings.

\section{Asymptotic fidelity of uniaxial transfer matrices}
\label{sec:pumping_uniaxial}
In this appendix, we show that uniaxial pumping protocols (class III) cannot distill entanglement from supply pairs in the interior $\mathscr{T}^\circ$ of the state-space tetrahedron, and we determine the attainable fidelity for supply pairs on its boundary. In the ordered basis $(e_t, e_\alpha, e_\beta, e_\gamma)$, to which all explicit vectors in this appendix refer, the transfer matrix of a uniaxial protocol is block diagonal,
\begin{align}
\label{eq:G_uniaxial}
    G = \frac{1}{2}\left[ M(1,g)_{(t,\alpha)} \oplus M(b,h)_{(\beta,\gamma)} \right] \eqdef \frac{1}{2}(A \oplus B).
\end{align}
The convergence argument based on the Perron-Frobenius theorem cannot be applied here, because $G$ is not a primitive matrix for a Class III protocol. Therefore, in this section, we construct an alternative convergence argument.

After $n$ rounds, the retained pair is described by $v_n = G^n v_0 = (t_n, \alpha_n, \beta_n, \gamma_n)^T$, and its fidelity is
\begin{align}
\label{eq:Fn_functional}
    F_n = \frac{\mathbf{1}^T v_n}{4\, t_n}, \qquad \mathbf{1} = (1,1,1,1)^T.
\end{align}
We assume the pumping condition $g,b,h>0$ and use the supply inequality $b+h \leq 1+g$ of Eq.~\eqref{eq:population_pg}, which is strict for supply pairs in $\mathscr{T}^\circ$ and an equality on the face $\mathscr{F}_g \subset \partial\mathscr{T}$ on which the supply population $p^0_g$ vanishes. For a coordinate vector $v$, we denote by $p_0(v) = F(v)$ and $p_\alpha(v)$, $p_\beta(v)$, $p_\gamma(v)$ the Bell populations of the normalized state according to Eq.~\eqref{eq:populations_expressed_in_coords}, where $p_\alpha$ belongs to the Bell state $\ket{B_\alpha}$ associated with the principal axis. For the supply pair, we write $p^0_\alpha = p_\alpha(v_0)$, etc.~Note the slight difference in notation here compared to the rest of the text. In this calculation, we need to differentiate between the populations of the supply state $p_k^0$ and the populations of the retained pair $p_k(v_n)$. In the main text, this distinction is not necessary and we omit the superscript $p^0_k = p_k$.
 
Since $G$ is block diagonal, it is reducible and hence not primitive. The Perron-Frobenius argument of Sec.~\ref{sec:definition_stabilizer_coordinates} therefore does not apply: the dominant eigenvalue of $G$ need not be simple, and the iteration may converge to a vector that depends on the initial state. We therefore do not use the right eigenvectors of $G$. Instead, we follow the two linear functionals that define the fidelity, the numerator $\mathbf{1}^T v$ and the trace $t$, which requires only the eigenvalues of the two blocks.
 
Every quantity we need is a linear functional $v \mapsto \ell^T v$, and under the iteration $\ell^T v_n = \ell^T G^n v_0$. In general, this mixes all components of $v_0$. If, however, $\ell$ is a left eigenvector of $G$, $\ell^T G = \lambda\, \ell^T$, then
\begin{align}
\label{eq:left_eigenvector_rescaling}
    \ell^T v_n = \lambda^n\, \ell^T v_0.
\end{align}
Because $G$ is block diagonal, a left eigenvector of one block, padded with zeros, is a left eigenvector of $G$. Both blocks are symmetric with constant diagonal, so $(1,1)$ and $(1,-1)$ are left eigenvectors of each. This yields the four left eigenvectors
\begin{align}
\label{eq:left_eigenvectors}
\begin{aligned}
    \ell_A  &= (1,\phantom{-}1,0,\phantom{-}0)^T, &\quad \ell_A^T G  &= \tfrac{1}{2}(1+g)\, \ell_A^T,  &\quad \ell_A^T v  &= 2t\,[p_0(v) + p_\alpha(v)],\\
    \ell_A' &= (1,-1,0,\phantom{-}0)^T,           &\quad (\ell_A')^T G &= \tfrac{1}{2}(1-g)\, (\ell_A')^T, &\quad (\ell_A')^T v &= 2t\,[p_\beta(v) + p_\gamma(v)],\\
    \ell_B  &= (0,\phantom{-}0,1,\phantom{-}1)^T, &\quad \ell_B^T G  &= \tfrac{1}{2}(b+h)\, \ell_B^T,  &\quad \ell_B^T v  &= 2t\,[p_0(v) - p_\alpha(v)],\\
    \ell_B' &= (0,\phantom{-}0,1,-1)^T,           &\quad (\ell_B')^T G &= \tfrac{1}{2}(b-h)\, (\ell_B')^T, &\quad (\ell_B')^T v &= 2t\,[p_\beta(v) - p_\gamma(v)],
\end{aligned}
\end{align}
where the subscripts refer to the principal block $A = M(1,g)_{(t,\alpha)}$ and the block $B = M(b,h)_{(\beta,\gamma)}$ that mixes the secondary axes, and the last column follows from Eq.~\eqref{eq:populations_expressed_in_coords}. Applied to the supply pair, the same relations express the eigenvalues through the supply populations, e.g.\ $\tfrac12(1+g) = p^0_0 + p^0_g$ and $\tfrac12(b+h) = p^0_0 - p^0_g$. The two functionals entering Eq.~\eqref{eq:Fn_functional} decompose as
\begin{align}
\label{eq:functional_split}
    \mathbf{1} = \ell_A + \ell_B, \qquad t = \tfrac{1}{2}\left(\ell_A + \ell_A'\right)^T v .
\end{align}
The fidelity numerator has a component in both blocks, whereas the trace lives entirely in the principal block.

\begin{proposition}
Let $g, b, h> 0$ with $b+h \leq 1 + g$, and define
\begin{align}
    \eta \defeq \frac{b + h}{1+g}, \quad R_n \defeq \frac{\ell_B^T v_n}{\ell^T_A v_n } = \frac{F_n - p_\alpha(v_n)}{F_n + p_\alpha(v_n)}.
\end{align}
Then for all $n \geq 0$
\begin{align}
    F_n \leq \tfrac{1}{2}(1 + \eta^n R_0).
\end{align}
In particular $\lim_{n \rightarrow \infty} F_n \leq \tfrac{1}{2}$ for every interior supply pair $v_0 \in \mathscr{T}^\circ$.
\end{proposition}
\begin{proof}
For every $n$, the vector $v_n/t_n$ corresponds to a physical state, so $\abs{\alpha_n /t_n} \leq 1$ because it is the expectation value of a Pauli string with eigenvalues $\pm 1$, i.e.~
\begin{align}
    \frac{\alpha_n}{t_n} = \langle S_\alpha \rangle_{\rho_n} = \Tr(S_\alpha \rho_n), \quad S_\alpha \in \{M, N, MN\}.
\end{align}
Since $t_n > 0$, we have $\alpha_n \leq t_n$ and eventually $(\ell'_A)^T v_n = t_n - \alpha_n \geq 0$. From the decomposition in Eq.~\eqref{eq:functional_split}, we can conclude the inequality
\begin{align}
\label{eq:bound_proof_uniaxial}
4 t_n = 2 \ell_A^T v_n + 2 (\ell_A')^T v_n \geq 2 \ell_A^T v_n =  2 \left[\tfrac{1}{2} (1+g) \right]^n (1 + \alpha_0) > 0.
\end{align}
With this, we can rewrite the expression for the fidelity $F_n$ to
\begin{align}
    F_n = \frac{(\ell_A + \ell_B)^T v_n}{4 t_n} \leq \frac{\ell_A^T v_n + \ell_B^T v_n}{2 \ell_A^T v_n} = \frac{1}{2}(1 + R_n).
\end{align}
Furthermore, we use the eigenvalue equations 
\begin{align*}
    \ell_A^T v_n &= \ell_A^T G^n v_0 = \left[\tfrac{1}{2}(1 + g) \right]^n \ell_A^Tv_0, \\
    \ell_B^T v_n &= \ell_B^T G^n v_0 = \left[\tfrac{1}{2}(b + h) \right]^n \ell_B^Tv_0,
\end{align*}
so that $R_n$ can be written in terms of $R_0$
\begin{align}
    R_n = \eta^n R_0,
\end{align}
and for $0 <\eta < 1$ the term $R_n$ contracts. Inserting $R_n = \eta^n R_0$ into the bound above proves the claim. For $v_0 \in \mathscr{T}^\circ$, the inequality~\eqref{eq:population_pg} is strict, so $0 < \eta < 1$ and $\eta^n R_0 \to 0$.
\end{proof}

Therefore, for interior points $v_0 \in \mathscr{T}^\circ$, uniaxial protocols are unable to improve the fidelity asymptotically. However, a supply state on the face $v_0 \in \mathscr{F}_g \subset \partial \mathscr{T}$ where $1 + g = b+h$ holds, shows no contractive dynamics $\eta^n =1$, leading to a possible entanglement distillation. In fact, by not using the approximation in Eq.~\eqref{eq:bound_proof_uniaxial}, one can obtain an exact expression for $F_n$. Define
\begin{align}
    \theta = \frac{1 - g}{1+g} \in [0,1), \quad  S_n = \frac{(\ell_A')^T v_n}{\ell_A^T v_n} = \frac{p_\beta(v_n) + p_\gamma(v_n)}{F_n + p_\alpha(v_n)}, \quad S_n = \theta^n S_0.
\end{align}
Using a similar calculation as in the proof, one obtains
\begin{align}
    F_n = \frac{1 + \eta^n R_0}{2(1 + \theta^n S_0)}.
\end{align}
In the case $v_0 \in \mathscr{F}_g$, we have $\eta = 1$, so inserting $R_0$ and $S_0$ yields
\begin{align}
    F_n = \frac{F_0}{F_0 + p_\alpha^0 + \theta^n(1 - F_0 - p^0_\alpha)} \xrightarrow{n \rightarrow \infty} \frac{F_0}{F_0 + p_\alpha^0}.
\end{align}

\section{Classification of transfer matrices}
\label{sec:classification_transfer_matrices}
The set of transfer matrices generated can be grouped into four classes, namely \textit{permutation protocols $(I)$}, \textit{replacement protocols $(II)$}, \textit{uniaxial pumping protocols $(III)$}, and \textit{biaxial pumping protocols $(IV)$}. Only class $IV$ protocols are proper distillation protocols in the sense that they work for all supply states (including interior states $v_o \in \mathscr{T}^\circ$) above a certain threshold for the supply fidelity $F_0$, while Class-III distillation only attains distillation dynamics for a specific subset of input states (see Appendix~\ref{sec:pumping_uniaxial}). Transfer matrices of type $I$ and $II$ do not show distillation properties at all and are only mentioned for completeness. 

To carry out this classification, we will not base the calculation on the Pauli string algebra but on an equivalent representation, namely the binary representation of the Pauli group \cite{calderbank1998, hostens2005, dehaene2003, jansen2022, nielsen2010}. With the stabilizers $M = s_M X_a X_b$ and $N = s_N Z_a Z_b$, any matched string $P$ supported on pair $j$ can be represented by a binary vector $\galois^2 = \{0, 1\}^{\times 2}$, e.g.~a string
\begin{align}
    P \defeq R_j(x) \defeq M^{x_m}_j N^{x_n}_j \quad \Longleftrightarrow \quad x \defeq (x_m, x_n) \in \galois^2,
\end{align}
where $R_j(x)$ is the shorthand notation for the Pauli string represented by $x$. The coordinate vector is then given by
\begin{align}
\label{eq:coordinates_binary}
    v_x \defeq \langle R_j(x) \rangle_\rho, \quad v_{(0,0)} = t, \; v_{(1, 0)} = \mu, \; v_{(0, 1)} = \nu, \; v_{(1, 1)} = \xi.
\end{align}
Note that the stabilizer group $S = \{I, M, N, MN\}$ is an abelian group, as for $A \neq B$, $AB = c_{AB} i C$, $c_{AB} = \pm 1$
\begin{align}
    (A \otimes A)(B \otimes B) = (AB) \otimes(AB) = -c_{AB}^2 C \otimes C = (BA) \otimes (BA) = (B \otimes B)(A \otimes A).
\end{align}
Therefore, the product of two strings amounts to the summation (modulo 2) of vectors in the two-element Galois field $\galois$, and we write $R_j(x) R_j(y) = R_j(x+y)$, because
\begin{align}
    R_j(x) R_j(y) = M_j^{x_m} \underbrace{N_j^{x_n} M_j^{y_m}}_{= M_j^{y_m} N_j^{x_n}} N_j^{y_n} = M^{x_m + y_m} N^{x_n + y_n} = R_j(x+y).
\end{align}
For each matched string $R_j(x)$, we can also define a partition among the two parties $A, B$ by $R_j(x) = s_M s_N W_{A,j}(x) W_{B,j}(x)$ where for any party $C$ we can write $W_C(x) = (X^{x_m} Z^{x_n})_{c_j}$ and $c_j$ is the qubit of party $C$ belonging to pair $j$. The 2-form $\omega(x,y) = x_m y_n + x_n y_m$ describes the commutation relation between two $W$-operators because
\begin{align}
\label{eq:commutation_binary_strings_W}
{
\begin{rcases}
    W(x) W(y) = X^{x_m}Z^{x_n} X^{y_m}Z^{y_n} = (-1)^{x_n y_m} W(x+y) \\
    W(y) W(x) = X^{y_m}Z^{y_n} X^{x_m}Z^{x_n} = (-1)^{x_m y_n} W(x+y)
\end{rcases}
}
 \; \Longrightarrow W(x) W(y) = (-1)^{\omega(x,y)} W(y) W(x),
\end{align}

For two-pair protocols, each matched string is given by $R_1(v) R_2(w)$. Let us define a vector space $V$ and two subspaces $V_1, V_2 \subset V$ by
\begin{align}
    V \defeq V_1 \oplus V_2 = \galois^2 \oplus \galois^2, \quad V_1 \defeq \galois^2 \oplus 0, \quad V_2 = 0 \oplus \galois^2.
\end{align}
where we write $R(x) = R_1(x_1) R_2(x_2)$ with $x_1 \in V_1, x_2 \in V_2$ being the components of vector $x = x_1 + x_2 \in V$. The binary vectors $x_j$ and the subspace $V_j$ represent the Pauli substring restricted to pair $j$.

As a next step, we analyze how we can describe transformations of $R_j(x)$ under conjugation $U^\dagger R_j(x) U$. First, we notice that for any two Pauli strings it holds that
\begin{align}
\label{eq:string_conjugation}
    e^{\pm i \tfrac{\pi}{4} P} Q e^{\mp i \tfrac{\pi}{4} P}  =
    \begin{cases}
        Q, &\text{if} \; [P, Q]=0, \\
        \pm i PQ, &\text{else}.
    \end{cases}
\end{align}
Partitioning $R(x)$ among the two sides of the setup (Alice and Bob) gives $R(x) = W(x)_A W(x)_B$ with $W(x)_C = (X^{x_1} Z^{x_2})_{c_1} (X^{x_3}Z^{x_4})_{c_2}$ for $x = (x_1, \ldots, x_4) \in \galois^4$ and $C \in \{A, B\}$. It is interesting to note that $W(x)^* = W(x)$. Moreover, let us write according to the Pauli string conjugation formula in Eq.~\eqref{eq:string_conjugation}
\begin{align}
\label{eq:conjugation_party_restriction}
    U^\dagger_A W(x)_A U_A = \eta(x) W_A( \Phi x),
\end{align}
where $\eta(x) \in \{\pm 1, \pm i\}$ and $\Phi: \galois^4 \rightarrow \galois^4$ describes the Pauli string multiplication tableau on $\galois^4$ according to the conjugation rule. Since we get for the mirrored side $B$
\begin{align}
    U_B^\dagger W(x)_B U_B \eqcomment{\tau} (U_A^*)^\dagger W(x)_A U_A^* = U_A^\dagger W(x)_A (U_A)^* \eqcomment{\tau} \eta(x)^* W(\Phi x)_B, 
\end{align}
where $\eqcomment{\tau}$ means equality under the mirror operator. Because $\eta(x)^* \eta(x) = \abs{\eta(x)}^2 = 1$ and $W(\Phi x)_A W(\Phi x)_B = R(\Phi x)$ we find for mirrored gates $U = U_A U_B = U_A \tau(U_A)^*$
\begin{align}
\label{eq:conjugation_binary}
    U^\dagger R(x) U = R(\Phi x).
\end{align}

We can define an extension of the 2-form $\omega$ on the new space $V$ by $\omega_V(x, y) = \omega(x_1, y_1) + \omega(x_2, y_2)$. The bilinear forms $\omega$ and $\omega_V$ are symplectic forms on $\galois^2$ and $\galois^4$, respectively, as bilinearity, alternation, and non-degeneracy are easily checked.

Furthermore, we can conclude that $\Phi$ is an element of the symplectic group $\Phi \in \Sp(4, \galois)$, meaning it leaves the symplectic form $\omega_V$ invariant. To show this, use the commutation relation $W(x) W(y) = (-1)^{\omega(x,y)} W(y) W(x)$ and plug in Eq.~\eqref{eq:conjugation_party_restriction} to obtain
\begin{align}
    \eta(x) \eta(y) W(\Phi x) W(\Phi y) = (-1)^{\omega_V(x,y)} \eta(x) \eta(y) W(\Phi y) W(\Phi x) \stackrel{\eqref{eq:commutation_binary_strings_W}}{=}(-1)^{\omega_V(\Phi x, \Phi y)} \eta(x) \eta(y) W(\Phi y) W(\Phi x),
\end{align}
so by comparing the sign exponents it immediately follows
\begin{align}
    \omega_V(\Phi x, \Phi y) = \omega_V(x, y) \quad \Longrightarrow \quad \Phi \in \Sp(4, \galois).
\end{align}
Moreover, $\Phi$ is injective. Suppose $\Phi x = \Phi y$. Then $R(\Phi x) - R(\Phi y) =  U^\dagger (R(x) - R(y)) U = 0$. Since $U$ is unitary, it can be inverted, and it follows that $R(x) -R(y) = 0 \Longrightarrow x = y$. It is interesting to note, that the map $\Phi$ can be further specified by Eq.~\eqref{eq:string_conjugation}, namely with $R(p) \defeq P$ and $R(q) = Q$ we can elegantly rewrite
\begin{align}
    \Phi(q) = q + \omega_V(q, p) \, p.
\end{align}

Within this framework, we can now develop a classification of transfer matrices. Recapitulate that the construction formula for the transfer matrix of protocol $(U, \Lambda_1)$ is
\begin{align*}
    \expval{C_2}_\Pi = \frac{1}{2}(\langle \widetilde{C}_2 + \widetilde{C}_2 \widetilde{\Lambda}_1 \rangle),
\end{align*}
where $\widetilde{O} = U^\dagger O U$. The operator $C_2$ is supported on pair 2, so $C_2 = R(w), w \in V_2$ and similarly $\Lambda_1 = s_\Lambda R(\lambda), \; \lambda \in V_1$. Using the rules for conjugation, we obtain
\begin{align}
    \widetilde{C}_2 = R(\Phi w), \quad \widetilde{\Lambda}_1 = s_\Lambda R(\ell), \quad  \widetilde{C}_2 \widetilde{\Lambda}_1 = s_\Lambda R(\Phi w) R(\ell) = s_\Lambda R(\Phi w + \ell),
\end{align}
where we introduced the vector $\ell \defeq \Phi \lambda = \ell_1 + \ell_2 \in V$ and define $L = \{0, \ell\}$. By substituting these expressions into the construction formula of the transfer matrices, the coordinates of the transfer matrix are then calculated for every $w \in V_2$ 
\begin{align*}
    v_w = \langle R(w)\rangle_{\rho_0 \otimes \rho} = \frac{1}{2} \sum_{x \in \Phi w + L} s(x) v_{x_1}^0 v_{x_2}^{\phantom{0}},
\end{align*}
with $s(x) = 1$ for $x =\Phi w$ and $s(x) = s_\Lambda$ for $x \neq \Phi w$. It is important to note that only one choice of $s_\Lambda$ is sensible: for a $\rho = \rho_0 = \dyad{\chi}{\chi}$, every coordinate is one and $2 v_w = 1 + s_\Lambda$, so a perfect state would be converted to the zero vector. Since we want to rule out such a behavior, we require $s_\Lambda = 1$. Thus the set of readout strings is reduced to $\{XX, -YY, ZZ\}$ (the signs are important) and we conclude
\begin{align}
\label{eq:transfer_matrix_binary_form}
    v_w = \langle R(w)\rangle_{\rho_0 \otimes \rho} = \frac{1}{2} \sum_{x \in \Phi w + L} v_{x_1}^0 v_{x_2}^{\phantom{0}}.
\end{align}
Define $E = E(\ell) = \langle \ell_1, \ell_2 \rangle = \{0, \ell_1, \ell_2, \ell\}$. If $\ell$ only has a non-zero component in either $V_1$ or $V_2$, then $E = L$ and $\ell$ is the only non-zero component of $E$, so
\begin{align}
    \dim(E) =
    \begin{cases}
        1, &\; \ell \in V_1 \cup V_2, \\
        2, &\; \ell \in V \setminus (V_1 \cup V_2).
    \end{cases}
\end{align}
$E$ is isotropic, i.e. $\omega_V(x,y) = 0 \; \forall x,y \in E$. Since $V_1 \perp V_2$ we have $\omega_V(\ell_1, \ell_2) = 0$ and $\omega_V(\ell, \ell_i) = \omega(\ell_1, \ell_i) + \omega(\ell_2, \ell_i) =0, \; i =1,2$. As we will see, the classification of transfer matrices reduces to the question whether $\Phi w \in E$ for $w \in V_2$. We discuss the two different dimensionalities of $E$ separately.

\textit{Case I -- $\dim(E) = 1$}. In this case, $\Phi w \in E$ implies $w = 0$. Suppose, $\Phi w \in E = \{0, \ell\}$. This would mean that $w \in \Phi^{-1}(E) = \{0, \lambda\}$. Because $w \in V_2 \cap \{0, \lambda\} = \{0\}$, the only possible solution is trivial.  

\textit{Case II -- $\dim(E) = 2$}. If $\dim(E) = 2$, both components of $\ell$ are non-zero $\ell_1, \ell_2 \neq 0$ and we can write
\begin{align}
\label{eq:alternative_description_E}
    E = \{v \in V: \omega(\ell_1, v_1) = 0 \land \omega(\ell_2, v_2) = 0 \},
\end{align}
which can be seen directly: to fulfill the first condition, $v_1$ must be either $0$ or $\ell_1$, and similarly for the second condition $v_2 \in \{0, \ell_2\}$. So $v \in \{0, \ell_1\} \oplus \{0, \ell_2\} = \{0, \ell_1, \ell_2, \ell_1 +\ell_2\} = E$. Also, because $\lambda \in V_1$ and $w \in V_2$ we find $\omega_V(\lambda, w) = 0$. Consequently, for any element $\Phi w \in \Phi(V_2)$ we find 
\begin{align}
\label{eq:ell_phi_w_relation}
    \omega_V(\ell, \Phi w) = \omega_V(\Phi \lambda, \Phi w) = \omega_V(\lambda, w) = 0,
\end{align}
and thus $\omega(\ell_1, [\Phi w]_1) = \omega(\ell_2, [\Phi w]_2)$. Define
\begin{align}
\label{eq:E_via_f}
   f: V_2 \rightarrow \galois, \; w \mapsto \omega_V(\ell_1, \Phi w) = \omega(\ell_1, [\Phi w]_1) \stackrel{\eqref{eq:ell_phi_w_relation}}{=} \omega(\ell_2, [\Phi w]_2).
\end{align}
To be precise, the function $f$ is a function of the Clifford operation $\Phi$, so $f \equiv f_\Phi$, but we omit an explicit reference in the notation of $f$. For vectors $w \in V_2$, both conditions in $\eqref{eq:alternative_description_E}$ are therefore identical and we can express the equivalence
\begin{align}
    w \in V_2: \quad \Phi w \in E \quad \Longleftrightarrow f(w) = 0.
\end{align}
So next we need to study the kernel of $f$. The dimension of the domain of $f$ is $\dim(V_2) = 2$ and $\image(f) \subseteq \galois \Rightarrow \dim(\image(f)) \leq 1$, so according to the rank-nullity theorem $\dim(\ker(f)) \in \{1, 2\}$. However, $f \equiv 0$ would mean that $\Phi(V_2) \subseteq E$, and because of the injectivity of $\Phi$ and $\abs{V_2} = \abs{E} = 4$, $\Phi(V_2) = E$. But since $\ell = \Phi \lambda \in E$ with $\lambda \in V_1$, there would be a $w_0 \in V_2$ with $\Phi(V_2) \ni\Phi w_0 = \ell \in E$. Due to the injectivity of $\Phi$ we find $w_0 = \lambda$ and $\lambda \in V_1 \cap V_2 = \{0\}$. Therefore we require $\dim(\ker(f))=1$ and $\ker(f) = \{0, \hat{\zeta}\}$ for some non-zero vector $\hat{\zeta} \in V_2$. $\Phi\hat{\zeta} \in E\setminus \{0, \ell\} = \{\ell_1, \ell_2\}$, because $\Phi \hat{\zeta} = 0$ would contradict the injectivity of $\Phi$ and $\Phi \hat{\zeta} = \ell = \Phi \lambda \Rightarrow \hat{\zeta} = \lambda \notin V_2$. Therefore, $\Phi \hat{\zeta} + L = \{\ell_1, \ell_2\}$ because $\ell_i + \ell = \ell_j$ for $i \neq j$.

Now we have developed the machinery to tackle different structures of transfer matrices, which we will order by the following classification.

\begin{proposition}[Classification of pumping protocols]
Pumping protocols can be classified into the following four classes. The representation strings of the coordinates in the primary pumping axis $\delta \leftarrow g \alpha$ (if existent) are $\delta \defeq  v_{\hat \zeta} = \langle R_2(\hat \zeta) \rangle$, $\alpha = v_{\ell_2}$ and for the pumping coordinate and $g \defeq v_{\ell_1}^0$ for the supply coordinate. The secondary axes $\beta, \gamma$ are represented by $\zeta', \zeta''$ so that $\beta = v_{\zeta'}$ and $\gamma = v_{\zeta''}$.

\underline{Class I (Permutation class).} A protocol is a permutation protocol if $\ell \in V_1$ and the transfer matrix is
\begin{align}
    G = \frac{1}{2}(1+ g) P,
\end{align}
where $P$ is a permutation among the coordinates $(\mu, \nu, \xi)$ and $g \defeq v_{\ell_1}^0$.

\underline{Class II (Replacement class).} A protocol is in the replacement class if $\ell \in V_2$. The coordinate transformation is $(w \in V_2)$
\begin{align}
    v' = \frac{1}{2}(t + v_{\ell_2}) \; v^0_{(\Phi w)_1}
\end{align}
i.e.~the non-zero elements on the columns $\{0, \ell_2\}$ in each row share the same value, so the transfer matrix is of rank $1$ and output coordinates are just a permutation of the supply coordinates.

\underline{Class III (Uniaxial pumping).} A protocol has uniaxial pumping dynamics if $\ell \in V \setminus (V_1 \cup V_2)$ and $\Phi (\ell_2) \in E \Longleftrightarrow 0 \neq \ell_2 \in \ker(f)$. Then $\hat{\zeta} = \ell_2$ and the transfer matrix reads
\begin{align}
    G = \frac{1}{2} \left[ M(1, g)_{(t, \alpha)} \oplus M(b, h)_{(\beta, \gamma)} \right], \quad
    M(x,y) \defeq
    \begin{pmatrix}
        x & y\\ y & x
    \end{pmatrix}.
\end{align}

\underline{Class IV (Biaxial pumping).} A protocol has biaxial pumping dynamics if $\ell \in V \setminus (V_1 \cup V_2)$ and $\Phi (\ell_2) \notin E \Longleftrightarrow 0 \neq \ell_2 \notin \ker(f)$. Then $\hat{\zeta} \neq \ell_2$ and
\begin{align}
    G = \frac{1}{2} P_{\alpha \delta} \left[ M(1, g)_{(t, \alpha)} \oplus M(b, h)_{(\beta, \gamma)} \right], \quad
    M(x,y) \defeq
    \begin{pmatrix}
        x & y\\ y & x
    \end{pmatrix}.
\end{align}
\end{proposition}
\begin{proof}
To (I): If $\ell \in V_1$ we have a one dimensional $E = \{0, \ell\}$ with $\ell_1 = \ell$ and $\ell_2 = 0$. So both elements in the set $\Phi w + L = \{\Phi w, \Phi w + \ell_1\}$ have the same component in the subspace $V_2$, so $v_{x_2}$ is identical in both terms of the sum of Eq.~\eqref{eq:transfer_matrix_binary_form}. Moreover, since for a $w \in V_2$ we have $\omega_V(\ell, \Phi w) = \omega_V(\Phi \lambda, \Phi w) = \omega_V(\lambda, w) = 0$, thus with $\ell = (\ell_1, 0)$ this reads $\omega(\ell_1, [\Phi w]_1) = 0$ and therefore $[\Phi w]_1 \in \{0, \ell_1\}$. To conclude, $\Phi w + L = \{([\Phi w]_1, [\Phi w]_2), ([\Phi w]_1 + \ell_1, [\Phi w]_2)\}$ and
\begin{align}
    v_w = \frac{1}{2}(v_0^0 + v^0_{\ell_1}) v_{(\Phi w)_2} \eqdef \frac{1}{2}(1 + g) v_{(\Phi w)_2}.
\end{align}
The map $w \mapsto (\Phi w)_2$ is an injective map, because $(\Phi w_0) = 0$ with $(\Phi w_0)_1 \in \{0, \ell_1\}$ implies $\Phi w \in \{0, \ell\} = E$ and $w \in {0, \lambda}$. With $w \in V_2$ follows $w=0$, hence the kernel of the map is trivial and the map is injective. An injective map can only permute the coordinates, so the statement follows.

To (II): The argument is analogous to the case (I). Here $\ell \in V_2$, so we also have a one-dimensional $E = \{0, \ell_2\}$ and the elements of the set $\Phi w + L = \{\Phi w, \Phi w + \ell_2\}$ have identical components on the $V_1$ subspace. By the same argument as before we find $(\Phi w)_2 \in \{0, \ell_2\}$ and hence
\begin{align}
    v_w = \frac{1}{2}(t + v_{\ell_2}) v^0_{(\Phi w)_1}\quad \Longrightarrow \quad  v \propto P v_0.
\end{align}
Also $w \mapsto (\Phi w)_1$ is an injective map, so $v^0_{(\Phi w)_1}$ includes all supply coordinates once, and thus the output vector is just the permuted supply vector $v_0$.

To (III) \& (IV): Let us treat the cases $\Phi w \in E$ and $\Phi w \notin E$ separately with $\dim(E)=2$ and $w \in V_2$. For $\Phi w \in E$, since the kernel of $f(w)$ is one-dimensional, we know that $w \in \{0, \hat \zeta\}$. We can calculate explicitly $\restr{\Phi w + L}{w=0} = \{0, \ell\}$ and also since $\Phi \hat{\zeta} \in \{\ell_1, \ell_2\} \Longrightarrow \restr{\Phi w + L}{w=\hat{\zeta}} = \{\ell_1, \ell_2\}$. With this, we can directly calculate the coordinate $v_w$:
\begin{align}
    w =0 \Longrightarrow \Phi w + L = \{0, \ell\}  &\Longrightarrow \quad v_0 = \frac{1}{2}(v^0_0 v_0 + v_{\ell_1}^0 v_{\ell_2}), \label{eq:w-succ_probability_eq} \\
    w =\hat\zeta \Longrightarrow \Phi w + L = \{\ell_1, \ell_2\}  &\Longrightarrow \quad v_{\hat{\zeta}} = \frac{1}{2}(v^0_{\ell_1} v_0+v_0^0v_{\ell_2}). \label{eq:w-primary_axis}
\end{align}
The first equation determines the acceptance probability, and the second equation is the update rule for the pumping axis. So $\Phi w \in E$ describes the dynamics of the primary coordinate (coordinate $\ell_2$) with fuel $g \defeq v_{\ell_1}$. If $\Phi (\ell_2) \in E$ (case (III)), it is immediately required that $\ell_2  \in \{0, \hat \zeta\} = \ker(f)$, so the only non-zero choice is $\ell_2 = \hat \zeta$ and the primary coordinate is pumped from itself, i.e.~$\alpha \leftarrow g \alpha$. By contrast, for case (IV), we have a pumping dynamic of the form $\alpha' \leftarrow g \alpha$ for $\alpha \neq \alpha'$.

Now we discuss the other case $\Phi w \notin E \Longleftrightarrow w \notin \ker(f)$. If $w$ is not in the kernel, we get $\omega(\ell_1, [\Phi w]_1) = \omega(\ell_2, [\Phi w]_2) = 1$. The symplectic form is non-zero when $(\Phi w)_1 \notin \{0, \ell_1\}$ and $(\Phi w)_2 \notin \{0, \ell_2\}$. Then the complements constitute the secondary supply coordinates $\{b,h\} \eqdef V_1 \setminus \{0, \ell_1\}$ and the secondary axes $\{\zeta', \zeta''\} \eqdef V_2 \setminus \{0, \ell_2\}$. We can then construct equations analogous to Eqs.~\eqref{eq:w-succ_probability_eq} \& \eqref{eq:w-primary_axis} and the transfer matrices follow.
\end{proof}

We may now proceed by translating the condition $\Phi \ell_2 \in E$ in terms of the protocol components $U$ and $\Lambda_1$. Since $\lambda \in V_1$ and $\ell_2 \in V_2$:
\begin{align}
    \omega_V(\ell,\Phi(\ell_2)) = \omega_V(\Phi \lambda, \Phi(\ell_2)) = \omega_V(\lambda, \ell_2) = 0.
\end{align}
Therefore, with the equivalence $\Phi(\ell_2) \in E \; \Longleftrightarrow \omega_V(\ell_2, \Phi(\ell_2))=0$ we find that the restrictions of $R(\ell_2)$ and $R(\Phi(\ell_2))$ on one of the parties $A, B$ must commute by Eq.~\eqref{eq:commutation_binary_strings_W}. Since $\ell = \Phi \lambda$, $\ell_2$ is the evolved readout operator $\Lambda_1$ restricted to the support of the second pair, so $\restr{\smash{\widetilde{\Lambda}_1}}{2}$. Analogously, $\Phi(\ell_2)$ is the evolved version of the previous operator, $R(\Phi(\ell_2)) = U^\dagger R(\ell_2) U$, and we can conclude
\begin{align}
    \Phi(\ell_2) \in E \quad \Longleftrightarrow \quad \left[\restr{\smash{\widetilde{\Lambda}_1}}{2,A}, \restr{U^\dagger \smash{\left( \restr{\smash{\widetilde{\Lambda}_1}}{2} \right)} U}{A} \right] = 0.
\end{align}
Here the restriction $\restr{\cdot}{2,A}$ is the simultaneous restriction to pair $2$ and party $A$, so effectively the Pauli matrix on qubit $a_2$.

\include{tex_supplement/classification}

\section{Bound on the attainable fidelity}
\label{sec:bound_F_infty}
In this section, we motivate our postulate that the coefficient $g$ of a biaxial transfer matrix must be minimal. We employ the binary vector formalism from Appendix~\ref{sec:classification_transfer_matrices}, where single qubit operations are represented as binary vectors by $Q = Q(x),\; x = (x_1, x_2) \in \galois^2$ with the one-sided embedding
\begin{align}
    Q(x) = i^{x_1 x_2} (X^{x_1} Z^{x_2})_A \otimes I_B.
\end{align}
We selected the subsystem $A$ for embedding without loss of generality. For two Pauli strings $R(x)$ and $Q(y)$, the commutation is determined by $W_A(x)$ of $R(x)$ and we find $R(x) Q(y) = (-1)^{\omega(x,y)} Q(y) R(x)$.  Let $\{E_j\}$ be a collection of single-qubit errors that transform Bell states into Bell states. First, we exploit the defining property of stabilizer group $S = \{I, M, N, MN\} \eqdef \{S_0, S_1, S_2, S_3\}$ and write $S_k = R(s_k)$, $E_j = Q(e_j)$ so that we obtain
\begin{align}
    \label{eq:stabilizer_and_error_commutation}
    E_j \ket{\chi} = E_j S_k \ket{\chi} = (-1)^{\omega(e_j, s_k)} S_k E_j \ket{\chi}.
\end{align}
Denote the Bell states by $\ket{B_k}$; then the state of a Bell-diagonal entangled pair can be written as $\rho = \sum_{k} \rho_k \dyad{B_k}{B_k}$ and any Bell state can be written as $\ket{B_k} = E_k \ket{\chi}$ for some single-qubit error $E_k$. Furthermore, we stipulate the convention $E_0 = I$ so $\ket{B_0} = \ket{\chi}$ and define the stabilizer coordinates $c_k \defeq \langle S_k \rangle$. For any Bell population we find
\begin{align}
     \label{eq:populations_expressed_in_coords}
    p_j &\defeq \mel{B_j}{\rho}{B_j} = \Tr{\dyad{B_j}{B_j} \rho} = \Tr{E_j \dyad{\chi}{\chi} E_j \rho} \nonumber \\
    &= \frac{1}{4}\sum_k \Tr{E_j S_k E_j \rho} =  \frac{1}{4} \sum_k (-1)^{\omega(e_j, s_k)} \langle S_k \rangle = \frac{1}{4} \sum_k (-1)^{\omega(e_j, s_k)} c_k.
\end{align}
We see that the Bell populations can be expressed as a sum over the stabilizer coordinates, where their sign is determined by the commutation relation $[E_j, S_k]$. An error $E_j$ with $j \neq 0$ commutes with exactly two elements of $S$, one of which is the identity. This gives rise to a sign pattern, which is responsible for the tetrahedral state space $\mathscr{T}$ in the stabilizer coordinates. For a pumping protocol, let $S_g$ be the stabilizer that is associated with the coefficient $g$. Then we find
\begin{align}
    \label{eq:population_pg}
    p_g = \frac{1}{4}( 1 + g - h -b ) \geq 0 \quad \Longleftrightarrow \quad 1 + g \geq b + h.
\end{align}
Consider a pumping protocol $\alpha' \leftarrow g \alpha$ with input coordinates $\alpha, \beta, \gamma$ and output coordinates $\alpha', \beta', \gamma'$. Denote the output fidelity after the $n$-th round by $F_n$ and the supply coordinates by $g, b, h$. The update equations are
\begin{align}
    t' = \frac{1}{2}(t + g \alpha), \quad \alpha' = \frac{1}{2}(gt + \alpha), \quad \beta' + \gamma' = \frac{1}{2}(b+h)(\beta + \gamma).
\end{align}
By inserting these equations into the fidelity expression, we can directly compute the one-step fidelity for a generic input pair $v$
\begin{align}
\label{eq:one_step_fidelity}
    4 F_1(v) -1 = \frac{\alpha + gt + (b+h)(\beta + \gamma)}{t + g \alpha},
\end{align}
and we obtain for a perfect input pair $v_\text{id} = (1, 1, 1, 1)^T$ after a single iteration
\begin{align}
    F_1(v_\text{id}) = \frac{1 + g + b+ h}{2(1 + g)}, \quad 1 -  F_1(v_\text{id}) = \frac{1 + g - b - h}{2(1 + g)} = \frac{p_{g}}{p_0 + p_{g}}.
\end{align}
We now proceed to prove the intuitive statement that the maximal one-step fidelity after a single step is upper bounded by $F_1(v_\text{id})$, i.e.
\begin{align}
    \max_{v \in \mathscr{T}} \; F_1(v) = F_1(v_\text{id}).
\end{align}
Using $p_\alpha \geq 0 \; \Leftrightarrow \; \beta + \gamma \leq 1 + \alpha$ on Eq.~\eqref{eq:one_step_fidelity} for a normalized input $v$ (i.e.~$t=1$) we find
\begin{align}
   4 F_1(v) -1 \leq \frac{\alpha + g + (b+h)(1 + \alpha)}{1 + g \alpha} \equiv f(\alpha),
\end{align}
with equality at $\beta + \gamma  = 1 + \alpha$. The right hand side now only depends on $\alpha$ and we can compute the derivative of $f(\alpha)$ directly:
\begin{align}
    (1 + g \alpha)^2 f'(\alpha) = (1 - g) \left[(1+g) + (b+h) \right] \geq 0,
\end{align}
with a strict inequality for $g < 1$. Consequently, $f(\alpha)$ is monotonically increasing on the domain $[-1, 1]$ and $f(\alpha) \leq f(1) = 4 F_1(v_\text{id})-1$ with equality iff $\alpha = 1$. At maximum we have $\alpha = 1$ that requires $\beta + \gamma = 2$, i.e.,~$\beta = \gamma = 1$, and therefore we can conclude
\begin{align}
    F_1(v) \leq \frac{1}{4}[1+ f(\alpha)] \leq \frac{1}{4}[1+ f(1)] = F_1(v_\text{id}).
\end{align}

Now consider the Perron eigenvector $v_\star$ of a transfer matrix. As $v_\star$ is a fixed point, we have $F_\infty(v_\star) = F_1(v_\star) \leq F_1(v_\text{id})$, and we can therefore determine a lower bound for the infidelity of the pumping protocol
\begin{align}
    1 - F_\infty \geq \frac{1 + g - b - h}{2(1 + g)}.
\end{align}
This result is used as an argument for our conjecture $g \le b \le h$ as optimal signature in the main text.
%

\section{Positivity of stabilizer coordinates}
Throughout the manuscript, we use the fact that an entangled pair above the pumping threshold $F > 1/2$ has positive coordinates.
\begin{align}
\label{eq:pos_coordinates}
    F > \frac{1}{2} \quad \Longrightarrow \quad \mu, \nu, \xi > 0.
\end{align}
In this section, we derive this implication explicitly. We start from the relation in Equation~\eqref{eq:stabilizer_and_error_commutation} which tells us that $\ket{B_j}$ is an eigenvector for every given stabilizer $S_i$
\begin{align}
    S_i \ket{B_j} = (-1)^{\omega(e_j, s_i)} \ket{B_j}.
\end{align}
Thus the spectral decomposition of $S_i$ is simply
\begin{align}
    S_i = \sum_j (-1)^{\omega(e_j, s_i)} \dyad{B_j}{B_j}.
\end{align}
Now consider an arbitrary but fixed coordinate $c_x \in \{\mu, \nu, \xi\}$ together with its associated stabilizer $S_x$ (e.g.~$c_x = \nu$ and $S_x = ZZ$ for $\chi = \phi^+$). Calculating $c_x$ directly from the definition of the coordinates yields
\begin{align}
\label{eq:coords_expressed_in_populations}
    c_x = \langle S_x \rangle_\rho = \Tr(S_x \rho) = \sum_j (-1)^{\omega(e_j, s_x)} p_j,
\end{align}
which is the inverse relation of Eq.~\eqref{eq:populations_expressed_in_coords}. The eigenvalue to the eigenvector $\ket{B_j}$ is only $+1$ for $j=0$ and $j=x$, and $-1$ otherwise. If we denote the other coordinates by $c_y$ and $c_z$, we can write
\begin{align}
    c_x = [p_0 + p_x] - [p_y + p_z].
\end{align}
The positivity of $c_x$ immediately follows from this expression:
\begin{align}
    c_x = [p_0 + p_x] - \underbrace{[p_y + p_z]}_{= 1 - [p_0 + p_x]} = 2 \underbrace{[p_0 + p_x]}_{\geq p_0 =F} -1 \geq 2F-1 > 0,
\end{align}
where the last inequality is forced by our pumping condition $F > \frac{1}{2}$.

Note that the positivity of coordinates does not imply $F> \tfrac{1}{2}$ and Eq.~\eqref{eq:pos_coordinates} is not an equivalence. As a counterexample, take $0 <\varepsilon \ll 1$, take $v= (\varepsilon, \varepsilon, \varepsilon)$ with $F(v) = \tfrac{1}{4}(1 + 3 \varepsilon) \approx \tfrac{1}{4}$.

\section{Pumping conditions for Class IV protocols}
\label{sec:pumping_conditions}
Here we show the proof for the pumping condition stated in the main text that $F_0 \geq \varphi^{-1} \approx 0.618$ implies $F_\infty \geq F_0$ for every Class IV protocol, where $\varphi = (1 + \sqrt{5})/2$ is the golden ratio. Subsequently, we demonstrate that a weaker bound suffices for protocols with optimal signatures (see Eq.~\eqref{eq:min_fuel_conjecture}).

We start from Eq.~\eqref{eq:transfer_matrix_structure} and select without loss of generality $\delta = \beta$. This gives the following coordinate relations
\begin{align}
    t' = \frac{1}{2}(t + g \alpha), \quad \beta' = \frac{1}{2}(gt + \alpha), \quad \alpha' = \frac{1}{2}(b \beta + h \gamma), \quad \gamma' = \frac{1}{2}(h \beta + b \gamma).
\end{align}
Since populations are linearly related to the stabilizer coordinates, we can formulate these equations in terms of populations of the retained pair $p_0, p_\alpha, p_\beta, p_\gamma$ and the supply pair $F_0, p_g^0, p_b^0, p_h^0$:
\begin{align}
    p_0' &= F_0 p_0 + p_g^0 p_\alpha, \label{eq:pumping_cond_pop1}\\
    p_\alpha' &= p_b^0 p_\beta + p_h^0 p_\gamma, \label{eq:pumping_cond_pop2}\\
    p_\beta' &= p_g^0 p_0 + F_0 p_\alpha, \label{eq:pumping_cond_pop3}\\
    p_\gamma' &= p_h^0 p_\beta + p_b^0 p_\gamma.\label{eq:pumping_cond_pop4}
\end{align}
Let $\pi = (p_0, p_\alpha, p_\beta, p_\gamma)^T$ be the population vector and $\mathcal{P}$ be the matrix representing the linear equations. In this picture, the fidelity is $F(\pi) = p_0 / \norm{\pi}_1$. Furthermore, define $m \defeq p_b^0 + p_h^0 = 1 - F_0 - p_g^0$. The linear equation gives rise to a Perron-Frobenius eigenvector $\pi_\star$ with the corresponding eigenvalue $\lambda_\star$
\begin{align}
\label{eq:eigenvalue_eq_populations}
    \mathcal{P} \pi_\star = \lambda_\star \pi_\star.
\end{align}

\begin{proposition}[Pumping threshold for Class IV protocols]
\label{prop:pumping_threshold}
If $F_0 \geq \varphi^{-1} = (\sqrt{5}-1)/2$, then $F_\infty \geq F_0$ with equality if and only if $m = p_b^0 = p_h^0 = 0$ for all Class IV protocols and their respective signatures.
\end{proposition}
\begin{proof}
We assume $F_0 > \tfrac{1}{2}$ and normalize the vector $\pi_\star$ by its zero-component $\pi_\star = (1, x, y, z)^T \geq 0$ so that $F_\infty = 1/ \norm{\pi_\star}_1$. The eigenvalue equation related to Eq.~\eqref{eq:pumping_cond_pop1} gives $\lambda_\star = F_0 + p_g^0 x \geq F_0 > m$. $F_0 > m$ since $F_0 - m = 2F_0 - 1 + p_g^0 > 0$ for $F_0 > \tfrac{1}{2}$. Summing the rows of Eq.~\eqref{eq:eigenvalue_eq_populations} gives
\begin{align}
\lambda_\star (1 + x + y+ z) = \sum_i (\lambda_\star \pi_\star)_i = \sum_i(\mathcal{P} \pi_\star)_i = (F_0 + p_g^0)(1+x) + \underbrace{(p_b^0 + p_h^0)}_{=m} (y + z).
\end{align}
Subtracting $m \norm{\pi_\star}_1$ from both sides and using $g = 2(F_0 + p_g^0)-1$ then gives
\begin{align}
\frac{\lambda_\star - m}{F_\infty} = g(1+x) \quad \Longleftrightarrow \quad F_\infty = \frac{F_0 + p_g^0 x - m}{g(1+x)}.
\end{align}
Because of that, we can calculate the difference $F_\infty - F_0$ directly
\begin{align}
\label{eq:threshold_difference}
    F_\infty - F_0 = \frac{(F_0 -m-F_0 g) + (p_g^0 - F_0 g)x}{g(1+x)} = \frac{(2 F_0 -1) [m - (F_0 + p_g^0 )x ]}{g(1+x)}.
\end{align}
where we used $g = 2(F_0 + p_g^0)-1$, $1 -g = 2m$ and $p_g^0 - F_0 g = -(2F_0 -1)(F_0 + p_g^0)$ in the last step. Since $2F_0 -1 > 0$ and $g(1+x)>0$ we find the equivalence
\begin{align}
\label{eq:threshold_equivalence1}
    F_\infty \geq F_0 \quad \Longleftrightarrow \quad (F_0 + p_g^0) x \leq m.
\end{align}
Now, let us consider Eq.~\eqref{eq:pumping_cond_pop4} for $\pi_\star$ giving $ \lambda_\star z =p_h^0 y + p_b^0 z$. Because $\lambda_\star - p_b^0 \geq m - p_b^0 = p_h^0$ we find that $z = p_h^0 y / (\lambda_\star - p_b^0) \leq y$. Through the eigenvalue equations, we obtain
\begin{align}
\label{eq:lambda_squared_x}
    \lambda_\star x = p_b^0 y + p_h^0z \leq m y, \quad \lambda_\star y = p_g^0 + F_0 x \quad \Longrightarrow \quad \lambda_\star^2 x  \leq \lambda_\star my = m (p_g^0 + F_0 x).
\end{align}
Earlier, we determined that $\lambda_\star \geq F_0$, so an insertion of this property in the previous inequality gives
\begin{align}
\label{eq:threshold_equivalence2}
    F_0^2 x \leq \lambda_\star^2 x \leq m(p_g^0 + F_0 x) \quad \Longrightarrow \quad x \leq \frac{m p_g^0}{F_0(F_0 - m)}.
\end{align}
With this relationship, we can further modify the equivalence in Eq.~\eqref{eq:threshold_equivalence1}:
\begin{align}
(F_0 + p_g^0) x \stackrel{\eqref{eq:threshold_equivalence2}}{\leq} m \frac{(F_0 + p_g^0) p_g^0}{F_0 (F_0 - m)},
\end{align}
so this bound is $\leq m$ iff
\begin{align}
\label{eq:p_g-squared_inequality}
    (F_0 + p_g^0) p_g^0 \leq F_0 (F_0 - m) = F_0 (2 F_0 - 1 + p_g^0) \quad \Longleftrightarrow \quad (p_g^0)^2 \leq F_0(2 F_0 -1).
\end{align}
Finally, plugging in the worst case estimation $p_g^0 \leq 1 -F_0$ yields
\begin{align}
 (1-F_0)^2 \leq F_0 (2 F_0 - 1) \quad \Longleftrightarrow \quad F_0^2 + F_0 -1 \geq 0 \quad \Longleftrightarrow \quad F_0 \geq \varphi^{-1}.
\end{align}
$\varphi^{-1} = (\sqrt{5}-1)/2 \approx 0.618$ is the inverse of the golden ratio and the positive root of the polynomial $F_0^2 + F_0 - 1$. To conclude, we established the following argument
\begin{align}
    F_0 \geq \varphi^{-1} \; \Longleftrightarrow \; (1-F_0)^2 \leq F_0 (2 F_0 - 1) \; \Longrightarrow \; (p_g^0)^2 \leq F_0(2 F_0 -1) \; \Longrightarrow \; (F_0 + p_g^0) x \leq m \; \Longleftrightarrow \; F_\infty \geq F_0.
\end{align}

The proposition follows, because for $m > 0$, we have $p_g^0 < 1 - F_0$ so the inequality in Eq.~\eqref{eq:p_g-squared_inequality} is strict and $F_\infty > F_0$. If $m=0$, Eq.~\eqref{eq:threshold_equivalence2} forces $x=0$, and Eq.~\eqref{eq:threshold_difference} gives $F_\infty=F_0$. Since we showed $m > 0 \Rightarrow F_\infty \neq F_0$ and have $p_b^0,p_h^0\geq0$, we conclude by contraposition that $F_\infty = F_0 \Rightarrow m=0$ and thus $F_\infty=F_0$ if and only if $p_b^0=p_h^0=0$.
\end{proof}

In fact, the derived pumping threshold is sharp, i.e.~for every $\tfrac{1}{2} < F_0< \varphi^{-1}$ there is a supply state with $F_\infty < F_0$. To show this, consider $\tfrac{1}{2} < F_0 < \varphi^{-1}$ with $p_h^0=0$, $p_b^0=\varepsilon$ and $p_g^0 = 1 -F_0 -\varepsilon$. Due to Eq.~\eqref{eq:pumping_cond_pop4} and the eigenvalue equation we have $\lambda_\star z = \varepsilon z \; \Longleftrightarrow \; (\lambda_\star - \varepsilon) z = 0$.
Since we already found $\lambda_\star > m = \varepsilon$ it forces $z=0$. Thus, Eq.~\eqref{eq:lambda_squared_x} is an identity $\lambda_\star^2 x = \varepsilon(p_g^0 + F_0 x)$, so $x = \varepsilon p_g^0/(\lambda_\star^2 - \varepsilon F_0)$. Moreover, by Eq.~\eqref{eq:threshold_equivalence2}
\begin{align}
    x \leq \frac{\varepsilon p_g^0}{F_0 (F_0 - \varepsilon)} \xrightarrow{\varepsilon \rightarrow 0} 0, \quad \lambda_\star = F_0 + p_g^0 x \xrightarrow{\varepsilon \rightarrow 0} F_0.
\end{align}

Inserting the identity for $x$ into the sign condition following from Eq.~\eqref{eq:threshold_difference}, we get
\begin{align}
\frac{(F_0 + p_g^0) x}{m} = \frac{(F_0 + p_g^0) p_g^0}{\lambda_\star^2 - \varepsilon F_0} \xrightarrow{\varepsilon \rightarrow 0} \frac{1 - F_0}{F_0^2} > 1 \quad \text{for } \tfrac{1}{2} < F_0 < \varphi^{-1},
\end{align}
so for small $\varepsilon$, the ratio exceeds 1 giving $F_\infty < F_0$ by Eq.~\eqref{eq:threshold_difference}. Therefore, we have shown that for every $\tfrac{1}{2} < F_0 < \varphi^{-1}$, there is a supply state with $F_\infty < F_0$, validating the sharpness of Proposition~\ref{prop:pumping_threshold}.

\begin{proposition}[Pumping threshold for ordered signatures]
For protocols with $p_b^0 \leq p_h^0$ and $p_b^0 + p_h^0 > 0$, specifically protocols with ordered signature $p_g^0 \leq p_b^0 \leq p_h^0$ the following holds
\begin{align}
    F_0 > \frac{1}{2} \quad \Longrightarrow \quad F_\infty > F_0.
\end{align}
\end{proposition}
\begin{proof}
The proof is similar to the procedure in the proof of Proposition~\ref{prop:pumping_threshold}. We begin by showing that the ordering implies $x \leq z \leq y$. From Eq.~\eqref{eq:pumping_cond_pop4} we get $(\lambda_\star - p_b^0) z = p_h^0 y$ and with $\lambda_\star - p_b^0 > m - p_b^0 = p_h^0 \geq 0$ it follows that $z \leq y$. The difference of the eigenvalue equation related to equations \eqref{eq:pumping_cond_pop2} and \eqref{eq:pumping_cond_pop4} is
\begin{align}
    \lambda_\star (x - z) = (p_b^0 - p_h^0) (y-z).
\end{align}
Since $p_b^0 \leq p_h^0$ and $y \geq z$, the right side is $\leq 0$, hence $x \leq z$. The sum of the second and fourth eigenvalue equations gives
 \begin{align}
     \lambda_\star(x +z) = m (y + z),
 \end{align}
so with the ordering $z \geq x$ we get
\begin{align}
    \lambda_\star x = my - (\lambda_\star -m) z \leq my - (\lambda_\star - m) x \quad \Longleftrightarrow \quad (2 \lambda_\star - m) x \leq my.
\end{align}
Multiplying the obtained inequality with $\lambda_\star$ and using the eigenvalue equation $\lambda_\star y = p_g^0 + F_0 x$ gives $\lambda_\star (2 \lambda_\star - m)x \leq m (p_g^0 + F_0 x)$ and with $\lambda_\star \geq F_0$ we find $F_0 (2F_0 - m) x \leq m (p_g^0 + F_0 x)$. Subtracting $mF_0 x$ on both sides yields
\begin{align}
    2 F_0(F_0 - m ) x \leq m p_g^0.
\end{align}
Thus, solving for $x$ and evaluating
\begin{align}
    (F_0 + p_g^0) x \leq m \underbrace{\frac{p_g^0 (F_0 + p_g^0)}{2 F_0 (F_0 - m)}}_{\eqdef Q}.
\end{align}
For $F_0 > \tfrac{1}{2}$ the fraction $Q$ is always smaller than one, because with $F_0  - m = 2 F_0 -1 +p_g^0$, we find
\begin{align}
    2 F_0 (F_0- m) - p_g^0(F_0 + p_g^0) = 2 F_0 (2 F_0 -1) + p_g^0 (F_0 - p_g^0) > 0,
\end{align}
with $F_0 > \tfrac{1}{2}$ and $p_g^0 \leq 1 -F_0<F_0$. Since $m > 0$ we can conclude that $(F_0 + p_g^0) x \leq m Q < m$, so we obtain the statement of the proposition $F_\infty > F_0$ by Eq.~\eqref{eq:threshold_difference}.
\end{proof}
%
%
%
%
\section{Full controllability of centrally controlled registers}
\label{sec:full_controllability_first_order}
Here we show that the star-shaped register of Sec.~\ref{sec:system_and_control_pump} is fully controllable, even in the presence of detuning errors and the approximation to the first-order Magnus expansion. We start from the Hamiltonian in Eq.~\eqref{eq:hamiltonian_secular_and_qubit_approx} where the secular approximation has been carried out. If the driving field is detuned relative to the central electron spin by $\Delta$, it acquires an additional term $\vec{\Omega}'(t) = \Omega_x(t) \hat{x} + \Omega_y(t) \hat{y} + \Delta \hat{z}$, which we will separate from the control Hamiltonian $\vec{\Omega}'(t) = \vec{\Omega}(t) + \Delta \hat{z}$ and get
\begin{align}
H'(t) = - \sum_j \vec{\omega}_j \cdot \vec{I}_j + m_s S_z\sum_j \vec{A}_j \cdot \vec{I}_j +\vec{\Omega}(t) \cdot \vec{S} + \Delta S_z,
\end{align}
with the $S_i$-operators representing the spin operators in the NV centers qubit approximation and $\vec{A}_j = A^j_\perp \hat{x}_j + A^j_{\para} \hat{z}_j$ with $\hat{z}_j \propto \vec{\omega}_j$ and $\omega_j = \vert \vec{\omega}_j \vert$. As already shown in the main text, we define an interaction picture
\begin{align}
    \partial_t U_I(t) = i U_I(t) (\vec{\Omega}(t) \cdot \vec{S} - \sum_j \omega_j I_z^j), \quad U_I(0) = I,
\end{align}
so that the Hamiltonian in this frame becomes
\begin{align}
    H_I(t) = (\vec{f}(t) \cdot \vec{S}) \Big[\sum_j\vec{A}_j'(t) \cdot \vec{I}_j + \Delta \Big].
\end{align} 
The term $(\vec{f}(t) \cdot \vec{S}) \Delta$ models the detuning due to off-resonant driving of the electron spin. We will call the function $\vec{f}(t)$ defined by $\vec{f}(t) \cdot \vec{S} \defeq U_I(t)^{\nodagger} S_z U_I^\dagger(t)$ the \textit{generalized modulation function}, which generalizes the scalar modulation (filter) functions \cite{cywinski2008a, biercuk2011b}. The vector $\vec{A}'(t) = A_\perp \cos(\omega_0t) \hat{x} + A_\perp \sin(\omega_0t) \hat{y} + A_\para \hat{z}$ represents the altered interaction due to the rotating frame w.r.t.~$-\omega_j I_z^j$. Without loss of generality, consider a pulse sequence of period $T$ that is resonant with spin $j = 0$ and suppresses the effective dynamics of all other spins. The average Hamiltonian over the period can be written as
\begin{align}
    \label{eq:hamiltonian_register_aht}
    \bar{H}_T = \frac{1}{T} \int_0^T H_I(t) \dd{t} = A_\perp \left[ (\vec{m}_R(\omega_0) \cdot \vec{S}) I_x + (\vec{m}_I(\omega_0) \cdot \vec{S})  I_y \right] +  (\vec{m}_R(0) \cdot \vec{S})\, (A_\para  I_z + \Delta),
\end{align}
with the filter function $\vec{m}(\omega)$ defined as the Fourier transform of the modulation function
\begin{align}
    \vec{m}(\omega) = \vec{m}_R(\omega) + i \vec{m}_I(\omega) =  \frac{1}{T} \int_0^T e^{i \omega t} \vec{f}(t) \dd{t}.
\end{align}
We now determine the controllability of the register from the dynamical Lie algebra generated by $\bar{H}_T$. As in standard dynamical-decoupling sequences, the spectral component at $\omega_0$ generates terms of the form $AX$ and $BY$ ($A, B \in \{X, Y, Z\}$), provided the filter function has weight at the nuclear frequency $\vec{m}(\omega_0) \neq 0$. Also noteworthy is that a Pauli string of the form $DZ$ is generated by the zero-frequency component. This DC component usually leads to a loss of robustness, since detuning errors from the control term will be translated to the effective Hamiltonian. Depending on the treatment of the component $\vec{m}(0)$, we will distinguish between two control problems.

\subsection{Zero-mean modulation}
\label{sec:zero_mean_mod}
To address these issues, the set of allowed controls is typically constrained to zero-mean modulations, $\langle \, \vec{f}(t) \, \rangle_T = 0$, as discussed in the main text. We will denote the set of control functions that fulfill this restriction for effective two-qubit gates by $\mathcal{C}_0$. In fact, almost all dynamical decoupling sequences that offer control error mitigation (e.g.~XY, AXY, CPMG, KDD, etc.) are elements of this control class. As we already noted from Eq.~\eqref{eq:hamiltonian_register_aht}, the  remaining terms are either of the form $AX$ or $BY$ and the corresponding DLA is generated by electron spin rotations and the AXY gates
\begin{align}
    \mathfrak{g}_0 = -i \langle XI, YI, ZX, ZY \rangle_\text{Lie} = -i \SpanR{} \{XI, YI, ZI, IZ, XX, XY, YX, YY, ZX, ZY\}.
\end{align}
Since the AXY gates $ZX, ZY$ generate the Lie algebra, any $U \in e^{\mathfrak{g}_0}$ can be generated by local operations and AXY-based two-qubit gates. Hence, restricting to $\pi$-pulse-based pulse sequences entails no loss of controllability within $\mathcal{C}_0$.

\subsection{Lifting the modulation condition}
\label{sec:non-zero_mean_mod}
Now, consider pulse sequences with non-zero mean modulation $\langle \, \vec{f}(t) \, \rangle_T \neq 0$, so that the Fourier transform contributes a DC-term $\vec{m}_R(0)$ in the effective Hamiltonian. The set of control functions associated with this condition will be denoted as $\mathcal{C}_*$. Lifting the control constraint from $\mathcal{C}_0$ to $\mathcal{C}_*$ yields full controllability $\mathfrak{g}_* \defeq \langle -i \bar{H}_T(f) \mid f \in \mathcal{C}_* \rangle_\text{Lie} \simeq \su(4)$ even in first order average Hamiltonian theory. The challenge, however, is to design a pulse sequence that leverages the expanded control space while simultaneously suppressing the detuning terms. One needs to impose an extra structure on the pulse sequences to cancel out the extra detuning term.

We start from Eq.~\eqref{eq:hamiltonian_register_aht} but rewrite it for the multi-spin case with additional detuning
\begin{equation}
\bar{H}=(\vec{m}_R(0)\cdot\vec{S})\,\Delta
+\sum_j\Big\{A_\perp^j\big[(\vec{m}_R(\omega_j)\cdot\vec{S})\,I_x^j+(\vec{m}_I(\omega_j)\cdot\vec{S})\,I_y^j\big]
+(\vec{m}_R(0)\cdot\vec{S})\,A_\parallel^j I_z^j\Big\}.
\end{equation}
If $\vec{m}(\omega)$ has an appreciable weight only at a single nuclear frequency $\omega_K$, the resulting gate acts selectively on spin $K$. We then consider two kinds of Hamiltonians: (i) a Hamiltonian generated with a zero-mean modulation function that generates a gate belonging to $e^{\mathfrak{g}_0}$ and is selective to a targeted spin $k$ (named $H_1$), and (ii) a Hamiltonian, denoted by $H_2$, that only consists of the (non-selective) $\vec{m}_R(0)$-term, which may be produced by an off resonant pulse sequence and $\expval*{\vec{f}(t)}_T \neq 0$ so that $\vec{m}(\omega_j) \approx 0 \, \forall j$:
\begin{align}
    H_1 &= a_x S_z I_x^k + a_y S_z I_y^k =: \vert \vec{a} \vert \, S_z (\hat{a} \cdot \vec{I}_K), \\
    H_2 &= \sum_j b^j S_x I_z^j + \Delta S_x.
\end{align}
While $H_1$ can be generated by standard AXY sequences, for example~\cite{casanova2015a}, an off-resonant AXY sequence with trailing $\pi/2$-pulses can generate $H_2$. To make use of the non-selective term, we need to suppress the control error terms that contain $\Delta$. The detuning term is canceled by the composite sequence
\begin{align}
    U = U_1^\dagger U_2^{\phantom{\dagger}} U_1^{\phantom{\dagger}} U_2^{\phantom{\dagger}}, \qquad U_l = e^{-i H_l t_l}
\end{align}
for evolution times $t_1,~t_2$ yielding a total period of $T = 2(t_1 + t_2)$. We will now proceed by analyzing the sequence with average Hamiltonian theory (AHT). Because $U_1$ is a finite $(\pi)$ rotation, first-order AHT is not accurate in the original frame. We therefore go to a toggling frame with respect to $H_\text{toggle}(t) = g(t) H_1$, where the toggling function $g(t)$ is equal to one during $H_1$ blocks and zero otherwise. Note that the toggled and original reference frames coincide at the end of the composite sequence. The average Hamiltonian of the composite sequence up to second order reduces to
\begin{align}
\label{eq:composite_sequence_aht}
    \bar{H}_\text{CS} = \frac{1}{T}(t_2 H_2 + t_2 \, \tilde{H}_2) - \frac{i}{2 T} t_2^2 [H_2, \tilde{H}_2] + \mathcal{O}(\norm{H_2}^3 T^2).
\end{align}
with $\tilde{H}_2 := U_1(t_1) H_2 U_1(t_1)^\dagger$ and $\partial_t U_1(t) = i U_1(t) H_\text{toggle}(t), \; U_1(0) = I$ being the transformation operation to the toggling frame. When $t_1$ is set so that it equals a $\pi$-rotation, i.e.~$U_1(t_1) \propto S_z \otimes (\hat{a}\cdot \vec{I}_K)$, we find $\tilde{H}_2$ to be
\begin{align}
    \tilde{H}_2 = b^k S_x I_z^k - \sum_{j \neq k} b^j S_x I_z^j - \Delta S_x.
\end{align}
Inserting to Eq.~\eqref{eq:composite_sequence_aht} leads to an effective Hamiltonian
\begin{align}
     \bar{H}_\text{CS} = \frac{2 t_2}{T} b^K S_x \otimes I_z^K.
\end{align}
This adds a linearly independent generator to the DLA while retaining the detuning robustness of its constituents. Since all nested commutators $[H_2, \ldots[H_2, \tilde{H_2}]]$ are zero, the series expansion is exact. No selectivity constraint is imposed on $H_2$, so it can be tuned to be as fast as possible for minimal gate times while inheriting the selectivity attributes from the standard AXY components.

By introducing this new elementary gate, we augment the dynamical Lie algebra (DLA) with an additional basis vector, $XZ$. The extension $\mathfrak{g}_* = \langle \mathfrak{g}_0, XZ \rangle_\text{Lie}$ of $\mathfrak{g}_0$ by an element $XZ \notin \mathfrak{g}_0$ generates a new algebra $\mathfrak{g}_* \simeq \su(4)$, as one can see by explicit construction of the missing Pauli strings:
\begin{align}
    [XZ, YI] \propto ZZ, \qquad &[XZ, XY] \propto IX,\\
    [XZ, ZI] \propto YZ, \qquad &[XZ, XX] \propto IY.
\end{align}
%
%
%
%
%

\end{document}

%% file: tex_supplement/classification.tex
{\small
\begin{longtable}{@{}l c c >{\raggedright\arraybackslash}p{0.5\textwidth}@{}}
  \caption{Class III -- uniaxial pumping: For a restriction to at most two gates in $U$ and setting $\chi = \phi^+$, there are 120 protocols divided into 12 subclasses. Each row collects all protocols sharing one transfer matrix. $(g, b, h)$ describes the coordinate roles of the supply state and $\hat{\zeta} \leftarrow g \zeta$ denotes the pumping axis structure $\hat \zeta \propto g + \zeta$. The first element in the column listing the member protocols is the readout string $\Lambda_1$.}\label{tab:subclass-III} \\ 
        \toprule
    Subclass & $2G$ & $(g,b,h)$ & Members $[P_1,P_2]$ \\ 
    \midrule
  \endfirsthead
  \multicolumn{4}{@{}l}{\itshape Table \thetable\ (continued)} \\ 
    \toprule
    Subclass & $2G$ & $(g,b,h)$ & Members $[P_1,P_2]$ \\ 
    \midrule
  \endhead
  \midrule
  \multicolumn{4}{r@{}}{\itshape continued on next page} \\ 
  \endfoot
  \bottomrule
  \endlastfoot
  III.1 & \raisebox{\dimexpr\ht\strutbox-\height\relax}{$\begin{pmatrix}1 & 0 & 0 & \xi_0 \\ 0 & \nu_0 & \mu_0 & 0 \\ 0 & \mu_0 & \nu_0 & 0 \\ \xi_0 & 0 & 0 & 1\end{pmatrix}$} & \makecell[t]{$(\xi_0,\,\nu_0,\,\mu_0)$ \\ \footnotesize $\xi \leftarrow \xi_0 \xi$} & \footnotesize $\phantom{-}XX\colon$\; $[XY,ZY]$, $[YI,XY]$, $[ZY,YI]$ \newline $-YY\colon$\; $[XI,XY]$, $[XY,XI]$ \newline $\phantom{-}ZZ\colon$\; $[II,XY]$, $[XY,II]$, $[XY,YY]$, $[YY,ZI]$, $[ZI,XY]$ \\ 
  \addlinespace[3pt]
  III.2 & \raisebox{\dimexpr\ht\strutbox-\height\relax}{$\begin{pmatrix}1 & 0 & 0 & \nu_0 \\ 0 & \xi_0 & \mu_0 & 0 \\ 0 & \mu_0 & \xi_0 & 0 \\ \nu_0 & 0 & 0 & 1\end{pmatrix}$} & \makecell[t]{$(\nu_0,\,\xi_0,\,\mu_0)$ \\ \footnotesize $\xi \leftarrow \nu_0 \xi$} & \footnotesize $\phantom{-}XX\colon$\; $[XY,YY]$, $[YY,ZI]$, $[ZI,XY]$ \newline $-YY\colon$\; $[II,XY]$, $[XY,II]$, $[XY,ZY]$, $[YI,XY]$, $[ZY,YI]$ \newline $\phantom{-}ZZ\colon$\; $[XI,XY]$, $[XY,XI]$ \\ 
  \addlinespace[3pt]
  III.3 & \raisebox{\dimexpr\ht\strutbox-\height\relax}{$\begin{pmatrix}1 & 0 & 0 & \xi_0 \\ 0 & \mu_0 & \nu_0 & 0 \\ 0 & \nu_0 & \mu_0 & 0 \\ \xi_0 & 0 & 0 & 1\end{pmatrix}$} & \makecell[t]{$(\xi_0,\,\mu_0,\,\nu_0)$ \\ \footnotesize $\xi \leftarrow \xi_0 \xi$} & \footnotesize $\phantom{-}XX\colon$\; $[II,ZY]$, $[XI,ZY]$, $[YY,XI]$, $[ZY,II]$, $[ZY,YY]$ \newline $-YY\colon$\; $[ZI,ZY]$, $[ZY,ZI]$ \newline $\phantom{-}ZZ\colon$\; $[XY,YI]$, $[YI,ZY]$, $[ZY,XY]$ \\ 
  \addlinespace[3pt]
  III.4 & \raisebox{\dimexpr\ht\strutbox-\height\relax}{$\begin{pmatrix}1 & 0 & 0 & \nu_0 \\ 0 & \mu_0 & \xi_0 & 0 \\ 0 & \xi_0 & \mu_0 & 0 \\ \nu_0 & 0 & 0 & 1\end{pmatrix}$} & \makecell[t]{$(\nu_0,\,\mu_0,\,\xi_0)$ \\ \footnotesize $\xi \leftarrow \nu_0 \xi$} & \footnotesize $\phantom{-}XX\colon$\; $[II,YY]$, $[XI,YY]$, $[YY,II]$, $[YY,ZY]$, $[ZY,XI]$ \newline $-YY\colon$\; $[XY,ZI]$, $[YY,XY]$, $[ZI,YY]$ \newline $\phantom{-}ZZ\colon$\; $[YI,YY]$, $[YY,YI]$ \\ 
  \addlinespace[3pt]
  III.5 & \raisebox{\dimexpr\ht\strutbox-\height\relax}{$\begin{pmatrix}1 & 0 & 0 & \mu_0 \\ 0 & \xi_0 & \nu_0 & 0 \\ 0 & \nu_0 & \xi_0 & 0 \\ \mu_0 & 0 & 0 & 1\end{pmatrix}$} & \makecell[t]{$(\mu_0,\,\xi_0,\,\nu_0)$ \\ \footnotesize $\xi \leftarrow \mu_0 \xi$} & \footnotesize $\phantom{-}XX\colon$\; $[ZI,ZY]$, $[ZY,ZI]$ \newline $-YY\colon$\; $[II,ZY]$, $[XY,YI]$, $[YI,ZY]$, $[ZY,II]$, $[ZY,XY]$ \newline $\phantom{-}ZZ\colon$\; $[XI,ZY]$, $[YY,XI]$, $[ZY,YY]$ \\ 
  \addlinespace[3pt]
  III.6 & \raisebox{\dimexpr\ht\strutbox-\height\relax}{$\begin{pmatrix}1 & 0 & 0 & \mu_0 \\ 0 & \nu_0 & \xi_0 & 0 \\ 0 & \xi_0 & \nu_0 & 0 \\ \mu_0 & 0 & 0 & 1\end{pmatrix}$} & \makecell[t]{$(\mu_0,\,\nu_0,\,\xi_0)$ \\ \footnotesize $\xi \leftarrow \mu_0 \xi$} & \footnotesize $\phantom{-}XX\colon$\; $[YI,YY]$, $[YY,YI]$ \newline $-YY\colon$\; $[XI,YY]$, $[YY,ZY]$, $[ZY,XI]$ \newline $\phantom{-}ZZ\colon$\; $[II,YY]$, $[XY,ZI]$, $[YY,II]$, $[YY,XY]$, $[ZI,YY]$ \\ 
  \addlinespace[3pt]
  III.7 & \raisebox{\dimexpr\ht\strutbox-\height\relax}{$\begin{pmatrix}1 & \xi_0 & 0 & 0 \\ \xi_0 & 1 & 0 & 0 \\ 0 & 0 & \nu_0 & \mu_0 \\ 0 & 0 & \mu_0 & \nu_0\end{pmatrix}$} & \makecell[t]{$(\xi_0,\,\nu_0,\,\mu_0)$ \\ \footnotesize $\mu \leftarrow \xi_0 \mu$} & \footnotesize $\phantom{-}XX\colon$\; $[XX,ZX]$, $[YI,XX]$, $[ZX,YI]$ \newline $-YY\colon$\; $[XI,XX]$, $[XX,XI]$ \newline $\phantom{-}ZZ\colon$\; $[II,XX]$, $[XX,II]$, $[XX,YX]$, $[YX,ZI]$, $[ZI,XX]$ \\ 
  \addlinespace[3pt]
  III.8 & \raisebox{\dimexpr\ht\strutbox-\height\relax}{$\begin{pmatrix}1 & \nu_0 & 0 & 0 \\ \nu_0 & 1 & 0 & 0 \\ 0 & 0 & \xi_0 & \mu_0 \\ 0 & 0 & \mu_0 & \xi_0\end{pmatrix}$} & \makecell[t]{$(\nu_0,\,\xi_0,\,\mu_0)$ \\ \footnotesize $\mu \leftarrow \nu_0 \mu$} & \footnotesize $\phantom{-}XX\colon$\; $[XX,YX]$, $[YX,ZI]$, $[ZI,XX]$ \newline $-YY\colon$\; $[II,XX]$, $[XX,II]$, $[XX,ZX]$, $[YI,XX]$, $[ZX,YI]$ \newline $\phantom{-}ZZ\colon$\; $[XI,XX]$, $[XX,XI]$ \\ 
  \addlinespace[3pt]
  III.9 & \raisebox{\dimexpr\ht\strutbox-\height\relax}{$\begin{pmatrix}1 & \xi_0 & 0 & 0 \\ \xi_0 & 1 & 0 & 0 \\ 0 & 0 & \mu_0 & \nu_0 \\ 0 & 0 & \nu_0 & \mu_0\end{pmatrix}$} & \makecell[t]{$(\xi_0,\,\mu_0,\,\nu_0)$ \\ \footnotesize $\mu \leftarrow \xi_0 \mu$} & \footnotesize $\phantom{-}XX\colon$\; $[II,ZX]$, $[XI,ZX]$, $[YX,XI]$, $[ZX,II]$, $[ZX,YX]$ \newline $-YY\colon$\; $[ZI,ZX]$, $[ZX,ZI]$ \newline $\phantom{-}ZZ\colon$\; $[XX,YI]$, $[YI,ZX]$, $[ZX,XX]$ \\ 
  \addlinespace[3pt]
  III.10 & \raisebox{\dimexpr\ht\strutbox-\height\relax}{$\begin{pmatrix}1 & \nu_0 & 0 & 0 \\ \nu_0 & 1 & 0 & 0 \\ 0 & 0 & \mu_0 & \xi_0 \\ 0 & 0 & \xi_0 & \mu_0\end{pmatrix}$} & \makecell[t]{$(\nu_0,\,\mu_0,\,\xi_0)$ \\ \footnotesize $\mu \leftarrow \nu_0 \mu$} & \footnotesize $\phantom{-}XX\colon$\; $[II,YX]$, $[XI,YX]$, $[YX,II]$, $[YX,ZX]$, $[ZX,XI]$ \newline $-YY\colon$\; $[XX,ZI]$, $[YX,XX]$, $[ZI,YX]$ \newline $\phantom{-}ZZ\colon$\; $[YI,YX]$, $[YX,YI]$ \\ 
  \addlinespace[3pt]
  III.11 & \raisebox{\dimexpr\ht\strutbox-\height\relax}{$\begin{pmatrix}1 & \mu_0 & 0 & 0 \\ \mu_0 & 1 & 0 & 0 \\ 0 & 0 & \xi_0 & \nu_0 \\ 0 & 0 & \nu_0 & \xi_0\end{pmatrix}$} & \makecell[t]{$(\mu_0,\,\xi_0,\,\nu_0)$ \\ \footnotesize $\mu \leftarrow \mu_0 \mu$} & \footnotesize $\phantom{-}XX\colon$\; $[ZI,ZX]$, $[ZX,ZI]$ \newline $-YY\colon$\; $[II,ZX]$, $[XX,YI]$, $[YI,ZX]$, $[ZX,II]$, $[ZX,XX]$ \newline $\phantom{-}ZZ\colon$\; $[XI,ZX]$, $[YX,XI]$, $[ZX,YX]$ \\ 
  \addlinespace[3pt]
  III.12 & \raisebox{\dimexpr\ht\strutbox-\height\relax}{$\begin{pmatrix}1 & \mu_0 & 0 & 0 \\ \mu_0 & 1 & 0 & 0 \\ 0 & 0 & \nu_0 & \xi_0 \\ 0 & 0 & \xi_0 & \nu_0\end{pmatrix}$} & \makecell[t]{$(\mu_0,\,\nu_0,\,\xi_0)$ \\ \footnotesize $\mu \leftarrow \mu_0 \mu$} & \footnotesize $\phantom{-}XX\colon$\; $[YI,YX]$, $[YX,YI]$ \newline $-YY\colon$\; $[XI,YX]$, $[YX,ZX]$, $[ZX,XI]$ \newline $\phantom{-}ZZ\colon$\; $[II,YX]$, $[XX,ZI]$, $[YX,II]$, $[YX,XX]$, $[ZI,YX]$ \\ 
\end{longtable}
}

{\small
\begin{longtable}{@{}l c c >{\raggedright\arraybackslash}p{0.35\textwidth}@{}}
  \caption{Class IV -- biaxial pumping: For a restriction to at most two gates in $U$ and setting $\chi = \phi^+$, there are 60 protocols divided into 24 subclasses. $(g, b, h)$ describes the coordinate roles of the supply state and $\hat{\zeta} \leftarrow g \zeta$ denotes the pumping axis structure $\hat \zeta \propto g + \zeta$. The first element in the column listing the member protocols is the readout string $\Lambda_1$.}\label{tab:subclass-IV} \\ 
\toprule
    Subclass & $2G$ & $(g,b,h)$ & Members $[P_1,P_2]$ \\ 
    \midrule
  \endfirsthead
  \multicolumn{4}{@{}l}{\itshape Table \thetable\ (continued)} \\ 
    \toprule
    Subclass & $2G$ & $(g,b,h)$ & Members $[P_1,P_2]$ \\ 
    \midrule
  \endhead
  \midrule
  \multicolumn{4}{r@{}}{\itshape continued on next page} \\ 
  \endfoot
  \bottomrule
  \endlastfoot
  IV.1 & \raisebox{\dimexpr\ht\strutbox-\height\relax}{$\begin{pmatrix}1 & 0 & 0 & \xi_0 \\ \xi_0 & 0 & 0 & 1 \\ 0 & \nu_0 & \mu_0 & 0 \\ 0 & \mu_0 & \nu_0 & 0\end{pmatrix}$} & \makecell[t]{$(\xi_0,\,\mu_0,\,\nu_0)$ \\ \footnotesize $\boldsymbol{\mu \leftarrow \xi_0 \xi}$} & \footnotesize $\phantom{-}XX\colon$\; $[IZ,ZY]$, $[ZX,IZ]$, $[ZY,ZX]$ \\ 
  \addlinespace[3pt]
  IV.2 & \raisebox{\dimexpr\ht\strutbox-\height\relax}{$\begin{pmatrix}1 & 0 & 0 & \nu_0 \\ \nu_0 & 0 & 0 & 1 \\ 0 & \xi_0 & \mu_0 & 0 \\ 0 & \mu_0 & \xi_0 & 0\end{pmatrix}$} & \makecell[t]{$(\nu_0,\,\mu_0,\,\xi_0)$ \\ \footnotesize $\boldsymbol{\mu \leftarrow \nu_0 \xi}$} & \footnotesize $\phantom{-}XX\colon$\; $[IZ,YY]$, $[YX,IZ]$, $[YY,YX]$ \\ 
  \addlinespace[3pt]
  IV.3 & \raisebox{\dimexpr\ht\strutbox-\height\relax}{$\begin{pmatrix}1 & 0 & 0 & \xi_0 \\ \xi_0 & 0 & 0 & 1 \\ 0 & \mu_0 & \nu_0 & 0 \\ 0 & \nu_0 & \mu_0 & 0\end{pmatrix}$} & \makecell[t]{$(\xi_0,\,\nu_0,\,\mu_0)$ \\ \footnotesize $\boldsymbol{\mu \leftarrow \xi_0 \xi}$} & \footnotesize $\phantom{-}ZZ\colon$\; $[IZ,XY]$, $[XX,IZ]$, $[XY,XX]$ \\ 
  \addlinespace[3pt]
  IV.4 & \raisebox{\dimexpr\ht\strutbox-\height\relax}{$\begin{pmatrix}1 & 0 & 0 & \nu_0 \\ \nu_0 & 0 & 0 & 1 \\ 0 & \mu_0 & \xi_0 & 0 \\ 0 & \xi_0 & \mu_0 & 0\end{pmatrix}$} & \makecell[t]{$(\nu_0,\,\xi_0,\,\mu_0)$ \\ \footnotesize $\boldsymbol{\mu \leftarrow \nu_0 \xi}$} & \footnotesize $-YY\colon$\; $[IZ,XY]$, $[XX,IZ]$, $[XY,XX]$ \\ 
  \addlinespace[3pt]
  IV.5 & \raisebox{\dimexpr\ht\strutbox-\height\relax}{$\begin{pmatrix}1 & 0 & 0 & \mu_0 \\ \mu_0 & 0 & 0 & 1 \\ 0 & \xi_0 & \nu_0 & 0 \\ 0 & \nu_0 & \xi_0 & 0\end{pmatrix}$} & \makecell[t]{$(\mu_0,\,\nu_0,\,\xi_0)$ \\ \footnotesize $\boldsymbol{\mu \leftarrow \mu_0 \xi}$} & \footnotesize $\phantom{-}ZZ\colon$\; $[IZ,YY]$, $[YX,IZ]$, $[YY,YX]$ \\ 
  \addlinespace[3pt]
  IV.6 & \raisebox{\dimexpr\ht\strutbox-\height\relax}{$\begin{pmatrix}1 & 0 & 0 & \mu_0 \\ \mu_0 & 0 & 0 & 1 \\ 0 & \nu_0 & \xi_0 & 0 \\ 0 & \xi_0 & \nu_0 & 0\end{pmatrix}$} & \makecell[t]{$(\mu_0,\,\xi_0,\,\nu_0)$ \\ \footnotesize $\boldsymbol{\mu \leftarrow \mu_0 \xi}$} & \footnotesize $-YY\colon$\; $[IZ,ZY]$, $[ZX,IZ]$, $[ZY,ZX]$ \\ 
  \addlinespace[3pt]
  IV.7 & \raisebox{\dimexpr\ht\strutbox-\height\relax}{$\begin{pmatrix}1 & \xi_0 & 0 & 0 \\ 0 & 0 & \nu_0 & \mu_0 \\ 0 & 0 & \mu_0 & \nu_0 \\ \xi_0 & 1 & 0 & 0\end{pmatrix}$} & \makecell[t]{$(\xi_0,\,\mu_0,\,\nu_0)$ \\ \footnotesize $\boldsymbol{\xi \leftarrow \xi_0 \mu}$} & \footnotesize $\phantom{-}XX\colon$\; $[IZ,ZX]$, $[ZX,ZY]$, $[ZY,IZ]$ \\ 
  \addlinespace[3pt]
  IV.8 & \raisebox{\dimexpr\ht\strutbox-\height\relax}{$\begin{pmatrix}1 & \nu_0 & 0 & 0 \\ 0 & 0 & \xi_0 & \mu_0 \\ 0 & 0 & \mu_0 & \xi_0 \\ \nu_0 & 1 & 0 & 0\end{pmatrix}$} & \makecell[t]{$(\nu_0,\,\mu_0,\,\xi_0)$ \\ \footnotesize $\boldsymbol{\xi \leftarrow \nu_0 \mu}$} & \footnotesize $\phantom{-}XX\colon$\; $[IZ,YX]$, $[YX,YY]$, $[YY,IZ]$ \\ 
  \addlinespace[3pt]
  IV.9 & \raisebox{\dimexpr\ht\strutbox-\height\relax}{$\begin{pmatrix}1 & \xi_0 & 0 & 0 \\ 0 & 0 & \mu_0 & \nu_0 \\ 0 & 0 & \nu_0 & \mu_0 \\ \xi_0 & 1 & 0 & 0\end{pmatrix}$} & \makecell[t]{$(\xi_0,\,\nu_0,\,\mu_0)$ \\ \footnotesize $\boldsymbol{\xi \leftarrow \xi_0 \mu}$} & \footnotesize $\phantom{-}ZZ\colon$\; $[IZ,XX]$, $[XX,XY]$, $[XY,IZ]$ \\ 
  \addlinespace[3pt]
  IV.10 & \raisebox{\dimexpr\ht\strutbox-\height\relax}{$\begin{pmatrix}1 & \nu_0 & 0 & 0 \\ 0 & 0 & \mu_0 & \xi_0 \\ 0 & 0 & \xi_0 & \mu_0 \\ \nu_0 & 1 & 0 & 0\end{pmatrix}$} & \makecell[t]{$(\nu_0,\,\xi_0,\,\mu_0)$ \\ \footnotesize $\boldsymbol{\xi \leftarrow \nu_0 \mu}$} & \footnotesize $-YY\colon$\; $[IZ,XX]$, $[XX,XY]$, $[XY,IZ]$ \\ 
  \addlinespace[3pt]
  IV.11 & \raisebox{\dimexpr\ht\strutbox-\height\relax}{$\begin{pmatrix}1 & \mu_0 & 0 & 0 \\ 0 & 0 & \xi_0 & \nu_0 \\ 0 & 0 & \nu_0 & \xi_0 \\ \mu_0 & 1 & 0 & 0\end{pmatrix}$} & \makecell[t]{$(\mu_0,\,\nu_0,\,\xi_0)$ \\ \footnotesize $\boldsymbol{\xi \leftarrow \mu_0 \mu}$} & \footnotesize $\phantom{-}ZZ\colon$\; $[IZ,YX]$, $[YX,YY]$, $[YY,IZ]$ \\ 
  \addlinespace[3pt]
  IV.12 & \raisebox{\dimexpr\ht\strutbox-\height\relax}{$\begin{pmatrix}1 & \mu_0 & 0 & 0 \\ 0 & 0 & \nu_0 & \xi_0 \\ 0 & 0 & \xi_0 & \nu_0 \\ \mu_0 & 1 & 0 & 0\end{pmatrix}$} & \makecell[t]{$(\mu_0,\,\xi_0,\,\nu_0)$ \\ \footnotesize $\boldsymbol{\xi \leftarrow \mu_0 \mu}$} & \footnotesize $-YY\colon$\; $[IZ,ZX]$, $[ZX,ZY]$, $[ZY,IZ]$ \\ 
  \addlinespace[3pt]
  IV.13 & \raisebox{\dimexpr\ht\strutbox-\height\relax}{$\begin{pmatrix}1 & 0 & 0 & \xi_0 \\ 0 & \nu_0 & \mu_0 & 0 \\ \xi_0 & 0 & 0 & 1 \\ 0 & \mu_0 & \nu_0 & 0\end{pmatrix}$} & \makecell[t]{$(\xi_0,\,\nu_0,\,\mu_0)$ \\ \footnotesize $\boldsymbol{\nu \leftarrow \xi_0 \xi}$} & \footnotesize $\phantom{-}ZZ\colon$\; $[XY,ZX]$, $[ZX,XY]$ \\ 
  \addlinespace[3pt]
  IV.14 & \raisebox{\dimexpr\ht\strutbox-\height\relax}{$\begin{pmatrix}1 & 0 & 0 & \nu_0 \\ 0 & \xi_0 & \mu_0 & 0 \\ \nu_0 & 0 & 0 & 1 \\ 0 & \mu_0 & \xi_0 & 0\end{pmatrix}$} & \makecell[t]{$(\nu_0,\,\xi_0,\,\mu_0)$ \\ \footnotesize $\boldsymbol{\nu \leftarrow \nu_0 \xi}$} & \footnotesize $-YY\colon$\; $[XY,YX]$, $[YX,XY]$ \\ 
  \addlinespace[3pt]
  IV.15 & \raisebox{\dimexpr\ht\strutbox-\height\relax}{$\begin{pmatrix}1 & 0 & 0 & \xi_0 \\ 0 & \mu_0 & \nu_0 & 0 \\ \xi_0 & 0 & 0 & 1 \\ 0 & \nu_0 & \mu_0 & 0\end{pmatrix}$} & \makecell[t]{$(\xi_0,\,\mu_0,\,\nu_0)$ \\ \footnotesize $\boldsymbol{\nu \leftarrow \xi_0 \xi}$} & \footnotesize $\phantom{-}XX\colon$\; $[XX,ZY]$, $[ZY,XX]$ \\ 
  \addlinespace[3pt]
  IV.16 & \raisebox{\dimexpr\ht\strutbox-\height\relax}{$\begin{pmatrix}1 & 0 & 0 & \nu_0 \\ 0 & \mu_0 & \xi_0 & 0 \\ \nu_0 & 0 & 0 & 1 \\ 0 & \xi_0 & \mu_0 & 0\end{pmatrix}$} & \makecell[t]{$(\nu_0,\,\mu_0,\,\xi_0)$ \\ \footnotesize $\boldsymbol{\nu \leftarrow \nu_0 \xi}$} & \footnotesize $\phantom{-}XX\colon$\; $[XX,YY]$, $[YY,XX]$ \\ 
  \addlinespace[3pt]
  IV.17 & \raisebox{\dimexpr\ht\strutbox-\height\relax}{$\begin{pmatrix}1 & 0 & 0 & \mu_0 \\ 0 & \xi_0 & \nu_0 & 0 \\ \mu_0 & 0 & 0 & 1 \\ 0 & \nu_0 & \xi_0 & 0\end{pmatrix}$} & \makecell[t]{$(\mu_0,\,\xi_0,\,\nu_0)$ \\ \footnotesize $\boldsymbol{\nu \leftarrow \mu_0 \xi}$} & \footnotesize $-YY\colon$\; $[YX,ZY]$, $[ZY,YX]$ \\ 
  \addlinespace[3pt]
  IV.18 & \raisebox{\dimexpr\ht\strutbox-\height\relax}{$\begin{pmatrix}1 & 0 & 0 & \mu_0 \\ 0 & \nu_0 & \xi_0 & 0 \\ \mu_0 & 0 & 0 & 1 \\ 0 & \xi_0 & \nu_0 & 0\end{pmatrix}$} & \makecell[t]{$(\mu_0,\,\nu_0,\,\xi_0)$ \\ \footnotesize $\boldsymbol{\nu \leftarrow \mu_0 \xi}$} & \footnotesize $\phantom{-}ZZ\colon$\; $[YY,ZX]$, $[ZX,YY]$ \\ 
  \addlinespace[3pt]
  IV.19 & \raisebox{\dimexpr\ht\strutbox-\height\relax}{$\begin{pmatrix}1 & \xi_0 & 0 & 0 \\ 0 & 0 & \nu_0 & \mu_0 \\ \xi_0 & 1 & 0 & 0 \\ 0 & 0 & \mu_0 & \nu_0\end{pmatrix}$} & \makecell[t]{$(\xi_0,\,\nu_0,\,\mu_0)$ \\ \footnotesize $\boldsymbol{\nu \leftarrow \xi_0 \mu}$} & \footnotesize $\phantom{-}ZZ\colon$\; $[XX,ZY]$, $[ZY,XX]$ \\ 
  \addlinespace[3pt]
  IV.20 & \raisebox{\dimexpr\ht\strutbox-\height\relax}{$\begin{pmatrix}1 & \nu_0 & 0 & 0 \\ 0 & 0 & \xi_0 & \mu_0 \\ \nu_0 & 1 & 0 & 0 \\ 0 & 0 & \mu_0 & \xi_0\end{pmatrix}$} & \makecell[t]{$(\nu_0,\,\xi_0,\,\mu_0)$ \\ \footnotesize $\boldsymbol{\nu \leftarrow \nu_0 \mu}$} & \footnotesize $-YY\colon$\; $[XX,YY]$, $[YY,XX]$ \\ 
  \addlinespace[3pt]
  IV.21 & \raisebox{\dimexpr\ht\strutbox-\height\relax}{$\begin{pmatrix}1 & \xi_0 & 0 & 0 \\ 0 & 0 & \mu_0 & \nu_0 \\ \xi_0 & 1 & 0 & 0 \\ 0 & 0 & \nu_0 & \mu_0\end{pmatrix}$} & \makecell[t]{$(\xi_0,\,\mu_0,\,\nu_0)$ \\ \footnotesize $\boldsymbol{\nu \leftarrow \xi_0 \mu}$} & \footnotesize $\phantom{-}XX\colon$\; $[XY,ZX]$, $[ZX,XY]$ \\ 
  \addlinespace[3pt]
  IV.22 & \raisebox{\dimexpr\ht\strutbox-\height\relax}{$\begin{pmatrix}1 & \nu_0 & 0 & 0 \\ 0 & 0 & \mu_0 & \xi_0 \\ \nu_0 & 1 & 0 & 0 \\ 0 & 0 & \xi_0 & \mu_0\end{pmatrix}$} & \makecell[t]{$(\nu_0,\,\mu_0,\,\xi_0)$ \\ \footnotesize $\boldsymbol{\nu \leftarrow \nu_0 \mu}$} & \footnotesize $\phantom{-}XX\colon$\; $[XY,YX]$, $[YX,XY]$ \\ 
  \addlinespace[3pt]
  IV.23 & \raisebox{\dimexpr\ht\strutbox-\height\relax}{$\begin{pmatrix}1 & \mu_0 & 0 & 0 \\ 0 & 0 & \xi_0 & \nu_0 \\ \mu_0 & 1 & 0 & 0 \\ 0 & 0 & \nu_0 & \xi_0\end{pmatrix}$} & \makecell[t]{$(\mu_0,\,\xi_0,\,\nu_0)$ \\ \footnotesize $\boldsymbol{\nu \leftarrow \mu_0 \mu}$} & \footnotesize $-YY\colon$\; $[YY,ZX]$, $[ZX,YY]$ \\ 
  \addlinespace[3pt]
  IV.24 & \raisebox{\dimexpr\ht\strutbox-\height\relax}{$\begin{pmatrix}1 & \mu_0 & 0 & 0 \\ 0 & 0 & \nu_0 & \xi_0 \\ \mu_0 & 1 & 0 & 0 \\ 0 & 0 & \xi_0 & \nu_0\end{pmatrix}$} & \makecell[t]{$(\mu_0,\,\nu_0,\,\xi_0)$ \\ \footnotesize $\boldsymbol{\nu \leftarrow \mu_0 \mu}$} & \footnotesize $\phantom{-}ZZ\colon$\; $[YX,ZY]$, $[ZY,YX]$ \\ 
\end{longtable}
}